\documentclass{giq6}
\usepackage{graphicx}

\renewcommand{\re}{\mathrm{e}}
\newcommand{\ri}{\mathrm{i}}
\newcommand{\rd}{\mathrm{d}}
\newtheorem{theorem}{Theorem}
\newtheorem{Prop}{Proposition}

\makeindex

\usepackage{amsmath,amssymb}
\newtheorem{remark}{Remark}%[section]
\def\bbbe\mathbb{E}
\def\bbbr{\mathbb{R}}

\def\bbbc{{\mathbb C}}
\def\bbbe{\mathbb{E}}
\def\bbbz{\mathbb{Z}}

\def\wedgecomma{\mathop{\wedge}\limits_{'}}
\def\ad{\mbox{ad}\,}
\def\diag{\mbox{diag}\,}

\def\Res{\mathop{\mbox{Res}\,}\limits}
\def\rank{\mbox{rank}\,}
\def\tr{\mbox{tr}\,}

\def\im{{\rm Im}\,}

\def\openone{\leavevmode\hbox{\small1\kern-3.3pt\normalsize1}}

\def\otimescomma{\mathop{\otimes}\limits_{'}}
\def\biglb{\big[\hspace*{-.7mm}\big[}
\def\bigrb{\big]\hspace*{-.7mm}\big]}

\def\Res{\mathop{\mbox{Res}\,}\limits}

\allowdisplaybreaks
\def\newpic#1{%
\def\emline##1##2##3##4##5##6{%
\put(##1,##2){\special{em:point #1##3}}%
\put(##4,##5){\special{em:point #1##6}}%
\special{em:line #1##3,#1##6}}}

\def\a{{\boldsymbol a}}
\def\b{{\boldsymbol b}}
\def\c{{\boldsymbol c}}
\def\d{{\boldsymbol d}}
\def\e{{\boldsymbol e}}
\def\T{{\boldsymbol T}}
\def\S{{\boldsymbol S}}
\def\R{{\boldsymbol R}}
\def\bPsi{{\boldsymbol \Psi}}
\def\bPhi{{\boldsymbol \Phi}}
\def\bepsilon{{\boldsymbol \epsilon}}

\def\q{{\boldsymbol q}}
\def\r{{\boldsymbol r}}

\def\det{\mbox{det\;}}
\def\openone{\leavevmode\hbox{\small1\kern-3.3pt\normalsize1}}

\newpic{}

\begin{document} % INITIALIZE - DONT CHANGE

\title[BASIC ASPECTS OF SOLITON THEORY]
{BASIC ASPECTS OF SOLITON THEORY}
\author{Vladimir S. Gerdjikov
%Second Author \MakeLowercase{and} Third Author
}
\address{Institute for Nuclear Research and Nuclear Energy, \\
Bulgarian Academy of Sciences,\\ 1784 Sofia, Bulgaria}

\begin{abstract}
This is a review of the main ideas of the inverse scattering
method (ISM) for solving nonlinear evolution equations (NLEE),
known as soliton equations. As a basic tool we use the fundamental
analytic solutions $\chi ^\pm(x,\lambda ) $ of the Lax operator
$L(\lambda)$. Then the inverse scattering problem for $L(\lambda)$
reduces to a Riemann-Hilbert problem. Such construction has been
applied to wide class of Lax operators, related to the simple Lie
algebras. We construct the kernel of the resolvent of $L(\lambda )
$ in terms of $\chi ^\pm(x,\lambda )$ and derive the spectral
decompositions of $L(\lambda ) $. Thus we can solve the relevant
classes of NLEE which include the NLS eq. and its
multi-component generalizations, the $N $-wave equations etc.
Applying the dressing method of Zakharov and Shabat we derive the
$N $-soliton solutions of these equations.

Next we explain that the ISM is a natural generalization of the
Fourier transform method. As appropriate generalizations of the
usual exponential function we use the so-called "squared
solutions" which are constructed again in terms of $\chi
^\pm(x,\lambda ) $ and the Cartan-Weyl basis of the relevant Lie
algebra. One can prove the completeness relations for the "squared
solutions" which in fact provide the spectral decompositions of
the recursion operator $\Lambda  $.

These  decompositions can be used to derive all fundamental
properties of the corresponding NLEE in terms of $\Lambda  $:
i)~the explicit form of the class of integrable NLEE; ii)~the
generating functionals of integrals of motion; iii)~the
hierarchies of Hamiltonian structures. We  outline the importance
of the classical $R $-matrices for extracting the involutive
integrals of motion.

\end{abstract}

\maketitle

{} {} {} {}

\begin{flushright}{}
%\vspace{-7mm}
%{\bf \today \qquad  F:/proceed/Varna.04/VSG/VSG-4.tex}
\end{flushright}

\section{Introduction} \label{intro}

The modern development of the soliton theory in the last three
decades of the 20-th century has lead to a number of important
applications and developments in several areas of contemporary
physics and mathematics, see \cite{ZMNP,FaTa,CaDeB,AblSe,APT}. In
this review I will outline the basic ideas of the inverse
scattering method (ISM) on the example of the nonlinear
Schr\"odinger equation (NLS) and its multicomponent
generalizations.

The integrability of the well known (scalar) NLS eq.:
\begin{equation}\label{eq:3.2}
\ri q_t + q_{xx} +2\epsilon |q(x,t)|^2 q(x,t) =0, \qquad \epsilon
=\pm 1 .
\end{equation}
was discovered by Zakharov and Shabat in their pioneer paper
\cite{ZS*71} which strongly stimulated the search of other
important integrable nonlinear evolution equations (NLEE). After
the Korteweg-de Vries equation, this was the second NLEE
integrable by the ISM. In the next few years the number of new
integrable NLEE was growing quickly: the modified KdV eq.
\cite{Wad}, the $N $-wave equations \cite{ZaMa,K,KRB}, the vector
NLS \cite{Ma1}, the Toda chain \cite{Ma2}, the principal chiral
field eq. \cite{ZaMik,Za*Mi} etc.

The simplest nontrivial multicomponent generalizations of NLS is
the vector NLS eq. known also as the Manakov model \cite{Ma1}:
\begin{equation}\label{eq:3.3}
\ri \vec{q}_t + \vec{q}_{xx} + 2 ({q}^\dag \vec{q}) \vec{q}(x,t)
=0,
\end{equation}
where $\vec{q} $ is an $n $-component complex-valued vector:
\[ \vec{q}(x,t) = \left( \begin{array}{c} q_1(x,t) \\ \vdots \\
q_n(x,t)
\end{array}\right), \]
tending to zero fast enough for $x\to\pm\infty  $. Another version
of the Manakov model is:
\begin{equation}\label{eq:3.3e}
\ri \vec{q}_t + \vec{q}_{xx} + 2 ({\vec{q}}{\,}^\dag\bepsilon
\vec{q}) \vec{q}(x,t) =0,
\end{equation}
where $\bepsilon =\diag ( \epsilon _1,\dots ,\epsilon_n )$ and
$\epsilon _n=\pm 1$. Both the scalar and vector NLS equations
(especially the one with $n=2 $) find wide applications in
nonlinear optics, plasma physics etc.

Equations (\ref{eq:3.3}) and (\ref{eq:3.3e}) are particular cases
of the matrix NLS eq. which is obtained from the system:
\begin{eqnarray}\label{eq:3.1}
\ri\q_t + \q_{xx} + 2 \q\r\q(x,t) =0,\nonumber\\
-\ri\r_t + \r_{xx} + 2 \r\q\r(x,t) =0.
\end{eqnarray}
with appropriate reductions (involution). Here  $\q(x,t) $ and
$\r^T(x,t) $ are $n\times m $-matrix-valued functions of $x $ and
$t $ with $n>1 $, $m>1 $ which are smooth enough and tend to zero
fast enough for $x\to\pm\infty  $. The best known involution
compatible with the evolution of (\ref{eq:3.1}) is
\begin{equation}\label{eq:3.5}
\r=B_- \q^\dag B_+^{-1}, \qquad B_\pm=\diag ( \eta_1^\pm,\dots
,\eta_m^\pm ), \qquad (\eta_s^\pm)^2=1.
\end{equation}
and the corresponding MNLS equation is of the form:
\begin{equation}\label{eq:3.6}
\ri\q_t + \q_{xx} + 2 \q B_- \q^\dag B_+^{-1} \q(x,t) =0
\end{equation}
For $n=m=1 $ and $r=\epsilon q^* $ the system (\ref{eq:3.1}) goes
into the scalar NLS equation; for $m=1$ and $n>1$ and with
appropriate choice of the involution (\ref{eq:3.5}) eq.
(\ref{eq:3.1}) can be transferred  into the Manakov model or into
eq. (\ref{eq:3.6}).

The MNLS (\ref{eq:3.6}) is known to be closely related to the
symmetric spaces \cite{ForKu*83}. All these versions of NLS are
solvable by applying the ISM. Their Lax representation $[L,M]=0 $
is provided by a generalization of the Zakharov-Shabat system:
\begin{eqnarray}\label{eq:4.1}
L\psi &\equiv & \left( \ri {\rd\over \rd x } + Q(x,t) - \lambda
J\right)
\psi (x,\lambda ) =0, \\
\label{eq:4.1M} M\psi &\equiv & \left(\ri {\rd \over \rd t} +
V_0(x,t) +\lambda V_1(x,t) -
2\lambda^2 J\right) \psi (x,\lambda ) =0, \\
Q(x,t) &=& \left( \begin{array}{cc} 0 & \q(x) \\ \r(x) & 0
\end{array} \right), \qquad J = \left( \begin{array}{cc} \openone
& 0 \\ 0 & -\openone \end{array} \right),
\end{eqnarray}
where $Q(x,t) $ and $J $ are $(n+m)\times (n+m) $ matrices with
compatible block structure and $V_0(x,t) $, $V_1(x,t) $ are
expressed in terms of $Q $ and its $x $-derivative:
\begin{equation}\label{eq:4.1V}
V_1(x,t) =2Q(x,t), \qquad V_0(x,t) = -[Q, \ad_J^{-1}Q] +
2\ri\ad_J^{-1}Q_x.
\end{equation}

\begin{remark}\label{rem:1}
One can consider Lax operators which are more general than
(\ref{eq:4.1}). For example, like in \cite{ZaMa,DrSok,G} one can
choose $Q(x,t) $ and $J $ to be elements of a simple Lie algebra
$\mathfrak{g} $ such that $J\in \mathfrak{h} $ and
$Q(x,t)=[J,\widetilde{Q}(x,t)] $. Such form of $Q(x,t) $ can
always be achieved by a gauge transformation commuting with $ J $.
Such $Q(x,t) $ span the co-adjoint orbit of $\mathfrak{g} $
passing through $J $ and can be viewed as the tangent plane to the
homogeneous space $\mathcal{G}/\mathcal{J} $. Here $\mathcal{G} $
is the Lie group with Lie algebra $\mathfrak{g} $ and $\mathcal{J}
$ is the subgroup of $\mathcal{G} $ commuting with $J $. Our
choice of $Q(x,t) $ in eq. (\ref{eq:4.1}) corresponds to the
symmetric space $SU(n+m)/S(U(n)\otimes U(m)) $, see also
\cite{ForKu*83}.

\end{remark}

\begin{remark}\label{rem:2}
An effective tool to impose involutions is the reduction group
introduced by  Mikhailov \cite{Mikh}. The involution
(\ref{eq:3.5}) (or $\bbbz_2$-reduction) can be written as:
\begin{equation}\label{eq:4.2}
B U^\dag (x,t,\lambda ^*) B^{-1} = U(x,t,\lambda ),
\end{equation}
where $B$ is an  automorphism of $\mathfrak{g} $ matrix such that
$B^2=\openone $, $[J,B]=0 $  and:
\begin{equation}\label{eq:11a}
U(x,t,\lambda ) = Q(x,t) -\lambda J, \qquad
B=\left(\begin{array}{cc} B_+ & 0 \\ 0 & B_- \end{array}\right).
\end{equation}
Reductions leading to new types of MNLS  systems are demonstrated
in \cite{varna04,manev04}.
\end{remark}

Section 2 is devoted to the direct scattering problem for the Lax
operator $L $. Our analysis is based on the notion of fundamental
analytic solution (FAS) which allows one to prove that the
scattering problem for $L $ is equivalent to a Riemann-Hilbert
problem (RHP) for FAS. Two minimal sets of scattering data
$\mathcal{T}_{1,2} $ are introduced and shown to determine
uniquely both the scattering matrix $T(\lambda ) $ and the
corresponding potential $Q(x) $.

In Section 3 we approach the solution of the inverse scattering
problem (ISP) through the RHP. The dressing Zakharov-Shabat method
\cite{ZaSha,ZaMik} for the symmetric spaces $SU(n+m)/S(U(n)\otimes
U(m)) $ is outlined and used to construct explicitly singular
solutions of the RHP. Thus we derive  reflectionless potentials
for $L $ and soliton solutions for the relevant NLEE. We also
define the resolvent of $L $ through the FAS  and prove the
completeness relation for the Jost solutions of $L $.

Section 4 starts with the Wronskian relations as a tool to study
the mapping between the potential $Q(x) $ and the scattering data
of $L $ generalizing the results of \cite{CaDe,CaDe2,AKNS,GKh1}.
Using them one is able to introduce the sets of `squared
solutions' $\{\bPsi \} $ and $\{\bPhi \} $ and prove that they are
complete in the space of all allowed potentials. This makes more
precise and explicit the results of \cite{TMF98}.  We derive the
expansions of $Q(x) $ and its variation $\ad_{J}^{-1}\delta Q(x) $
over $\{\bPsi \} $ and $\{\bPhi \} $ and demonstrate that the
elements of the minimal sets of scattering data $\mathcal{T}_{1,2}
$ and their variations appear as expansion coefficients. We also
introduce the generating operators $\Lambda _\pm $ for which
$\{\bPsi \} $ and $\{\bPhi \} $ are sets of eigen- and adjoint
functions.

The tools developed in Section 4 are used in Section 5 to describe
the fundamental properties of the NLEE. There we prove a theorem
stating the equivalence of the NLEE to a corresponding set of
linear evolution equations for the scattering data. Next we derive
the hierarchy of the integrals of motion from the principle series
in terms of $Q $ and show that they play the role of Hamiltonians
for the MNLS type equations. We display the hierarchy of
Hamiltonian structures for these NLEE. In short we have
demonstrated the complete analogy between the ISM and the usual
Fourier transform thus generalizing the results of
\cite{AKNS,DJK1,KN79,G,vg-ya,GYa}.

Using the method of the classical $R $-matrix \cite{FaTa} we
derive the Poisson brackets between the elements of the scattering
matrix. As a consequence we prove that the integrals of motion
from the principle series are in involution.

We end by a brief discussion in Section 6 on related methods and
topics.

\section{Direct and inverse scattering problems for $L $ }\label{sec:2}
\subsection{The scattering problem for $L $}\label{ssec:2.1}

Here we briefly outline the scattering problem for the system
(\ref{eq:4.1}) for the class of potentials $Q(x,t) $ satisfying
the following:

{\bf Condition C1:}  $Q(x,t) $ are smooth enough and fall off to
zero fast enough for $x\to\pm\infty  $ for all $t $.

{\bf Condition C2:}  $Q(x,t) $ is such that $L $ has a finite
number of simple eigenvalues $\lambda _j^\pm\in \bbbc_\pm $ for
all $t $.

In this subsection $t $  plays the role of an additional
parameter; for the sake of brevity the $t $-dependence is not
always shown. Condition C2 can not be formulated as a set of
explicit conditions on $Q(x,t) $; its precise meaning will become
clear below. The main tool here are the Jost solutions defined by
their asymptotics at $x\to\pm\infty  $:
\begin{equation}\label{eq:5.1}
\lim_{x\to\infty } \psi (x,\lambda )\re^{\ri\lambda Jx} =\openone
, \qquad \lim_{x\to -\infty } \phi (x,\lambda )\re^{\ri\lambda Jx}
=\openone ,
\end{equation}

Along with the Jost solutions we introduce
\begin{equation}\label{eq:5.1a}
\xi(x,\lambda ) =\psi (x,\lambda )\re^{\ri\lambda Jx}, \qquad
\varphi (x,\lambda ) =\phi (x,\lambda )\re^{\ri\lambda Jx};
\end{equation}
which satisfy the following linear integral equations
\begin{eqnarray}\label{eq:5.2}
\xi(x,\lambda ) &=& \openone + \ri \int_{\infty }^{x} \rd y
\re^{-\ri\lambda J(x-y)} Q(y) \xi(y,\lambda ) \re^{\ri\lambda
J(x-y)},
\\
\label{eq:5.2'} \varphi (x,\lambda ) &=& \openone + \ri
\int_{-\infty }^{x} \rd y \re^{-\ri\lambda J(x-y)} Q(y) \varphi
(y,\lambda ) \re^{\ri\lambda J(x-y)}.
\end{eqnarray}

These are Volterra type equations which, as is well known always
have solutions providing one can ensure the convergence of the
integrals in the right hand side. For $\lambda  $ real the
exponential factors in (\ref{eq:5.2}) and (\ref{eq:5.2'}) are just
oscillating and the convergence is ensured by condition C1.

Obviously the Jost solutions as whole can not be extended for $\im
\lambda \neq 0 $. However some of their columns can be extended
for $\lambda \in \bbbc_+ $, others -- for $\lambda \in \bbbc_- $.
Indeed, the equation (\ref{eq:5.2}) for the first column of
$\xi(x,\lambda ) $ contains only the exponential factor
$\re^{\ri\lambda(x-y) } $ which falls off for $\im \lambda <0 $.
More precisely we can write down the Jost solutions $\psi
(x,\lambda ) $ and $\phi (x,\lambda ) $ in the following
block-matrix form:
\begin{equation}\label{eq:5.3}
\psi (x,\lambda ) = \left(|\psi^- (x,\lambda )\rangle , |\psi^+
(x,\lambda )\rangle \right), \qquad  \phi (x,\lambda ) =
\left(|\phi^+ (x,\lambda ) \rangle , |\phi^- (x,\lambda )\rangle
\right),
\end{equation}
where the superscript $+ $ and (resp. $- $) shows that the
corresponding block-matrix allows analytic extension for $\lambda
\in \bbbc_+ $ (resp. $\lambda \in \bbbc_- $).

Solving the direct scattering problem means given the potential
$Q(x) $ to find the scattering matrix $T(\lambda ) $. By
definition $T(\lambda ) $ relates the two Jost solutions:
\begin{equation}\label{eq:6.1}
\phi (x,\lambda ) =\psi (x,\lambda )T(\lambda ), \qquad T(\lambda
)= \left(\begin{array}{cc} \a^+(\lambda ) & -\b^-(\lambda )
\\
\b^+(\lambda ) & \a^-(\lambda ) \end{array}\right)
\end{equation}
and has compatible block-matrix structure.  In what follows we
will need also the inverse of the scattering matrix:
\begin{equation}\label{eq:6.2i}
\psi (x,\lambda ) =\phi (x,\lambda )\hat{T}(\lambda ), \qquad
\hat{T}(\lambda )\equiv \left(\begin{array}{cc} \c^-(\lambda ) &
\d^-(\lambda ) \\ -\d^+(\lambda ) & \c^+(\lambda )
\end{array}\right),
\end{equation}
where
\begin{subequations}\label{eq:6.2}
\begin{eqnarray}\label{eq:6.2-cm}
\c^-(\lambda ) &=& \hat{\a}^+(\lambda ) (\openone +\rho ^-\rho
^+)^{-1} = (\openone +\tau^+\tau^-)^{-1} \hat{\a}^+(\lambda ), \\
\label{eq:6.2-dm} \d^-(\lambda ) &=& \hat{\a}^+(\lambda )\rho ^-
(\lambda ) (\openone  +\rho ^+\rho ^-)^{-1} = (\openone +\tau^+
\tau^-)^{-1} \tau^+ (\lambda ) \hat{\a}^-(\lambda ), \\
\label{eq:6.2-cp} \c^+(\lambda ) &=& \hat{\a}^-(\lambda )
(\openone +\rho^+\rho ^-)^{-1}= (\openone +\tau^-
\tau^+)^{-1}\hat{\a}^-(\lambda ), \\
\label{eq:6.2-dp} \d^+(\lambda ) &=& \hat{\a}^-(\lambda )\rho
^+(\lambda ) (\openone  +\rho ^-\rho ^+)^{-1}= (\openone
+\tau^-\tau^+)^{-1} \tau^- (\lambda )\hat{\a}^+(\lambda ).
\end{eqnarray}
\end{subequations}
The diagonal blocks of both $T(\lambda ) $ and $\hat{T}(\lambda )
$  allow analytic continuation off the real axis, namely
$\a^+(\lambda ) $, $\c^+(\lambda ) $ are analytic functions of
$\lambda  $ for $\lambda \in \bbbc_\pm $, while  $\a^-(\lambda )
$, $\c^-(\lambda ) $  are analytic functions of $\lambda$ for
$\lambda \in \bbbc_\pm $.

By $\rho ^\pm(\lambda ) $ and $\tau^\pm(\lambda ) $ above we have
denoted the multicomponent generalizations of the reflection
coefficients (for the scalar case, see
\cite{AKNS,CaDe,GKh1,KN79}):
\begin{equation}\label{eq:rho-tau}
\rho ^\pm(\lambda ) =\b^\pm\hat{\a}^\pm (\lambda ) =\hat{\c}^\pm
\d^\pm(\lambda ), \qquad \tau^\pm(\lambda ) =\hat{\a}^\pm
\b^\mp(\lambda ) =\d^\mp\hat{\c}^\pm (\lambda ) ,
\end{equation}
We will need also the asymptotics  for $\lambda \to\infty  $:
\begin{eqnarray}\label{eq:6.2b}
&& \lim_{\lambda \to -\infty } \phi (x,\lambda ) \re^{\ri\lambda Jx}
= \lim_{\lambda \to\infty } \psi (x,\lambda ) \re^{\ri\lambda Jx}
=\openone ,
\qquad \lim_{\lambda \to\infty } T(\lambda )  =\openone ,\\
&& \lim_{\lambda \to\infty } \a^+(\lambda )  = \lim_{\lambda
\to\infty } \c^-(\lambda )  = \openone , \qquad \lim_{\lambda
\to\infty } \a^-(\lambda )  = \lim_{\lambda \to\infty }
\c^+(\lambda )  = \openone . \nonumber
\end{eqnarray}

The inverse to the Jost solutions $\hat{\psi }(x,\lambda) $ and
$\hat{\phi }(x,\lambda) $ are solutions to:
\begin{equation}\label{eq:L-inv}
\ri {\rd\hat{\psi }\over \rd x } - \hat{\psi }(x,\lambda)
(Q(x)-\lambda J) =0,
\end{equation}
satisfying the conditions:
\begin{equation}\label{eq:J-inv}
\lim_{x\to\infty } \re^{-\ri\lambda Jx}\hat{\psi
}(x,\lambda)=\openone , \qquad \lim_{x\to -\infty } \re^{-\ri\lambda
Jx}\hat{\phi }(x,\lambda)=\openone .
\end{equation}

Now it is the collections of rows of $\hat{\psi }(x,\lambda) $ and
$\hat{\phi }(x,\lambda) $ that possess analytic properties in
$\lambda  $:
\begin{equation}\label{eq:psi-inv}
\hat{\psi }(x,\lambda) = \left( \begin{array}{c} \langle \hat{\psi
}^+(x,\lambda) | \\ \langle \hat{\psi }^-(x,\lambda) |
\end{array} \right), \qquad
\hat{\phi }(x,\lambda) = \left( \begin{array}{c} \langle \hat{\phi
}^-(x,\lambda) | \\ \langle \hat{\phi }^+(x,\lambda) |
\end{array} \right),
\end{equation}
Just like the Jost solutions, their inverse (\ref{eq:psi-inv}) are
solutions to linear equations (\ref{eq:L-inv}) with regular
boundary conditions (\ref{eq:J-inv}); therefore they can have no
singularities in their regions of analyticity. The same holds true
also for the scattering matrix $T(\lambda )=\hat{\psi }(x,\lambda)
\phi (x,\lambda) $ and its inverse $\hat{T}(\lambda )=\hat{\phi
}(x,\lambda) \psi (x,\lambda) $, i.e.
\begin{equation}\label{eq:a-pm}
\a^+(\lambda ) = \langle \hat{\psi }^+(x,\lambda) |\phi^+
(x,\lambda) \rangle , \qquad \a^-(\lambda ) = \langle \hat{\psi
}^-(x,\lambda) |\phi^- (x,\lambda) \rangle ,
\end{equation}
as well as
\begin{equation}\label{eq:c-pm}
\c^+(\lambda ) = \langle \hat{\phi }^+(x,\lambda) |\psi^+
(x,\lambda) \rangle , \qquad \c^-(\lambda ) = \langle \hat{\phi
}^-(x,\lambda) |\psi^- (x,\lambda) \rangle ,
\end{equation}
are analytic for $\lambda \in \bbbc_\pm $ and have no
singularities in their regions of analyticity.  However they may
become degenerate (i.e., their determinants may vanish) for some
values $\lambda _j^\pm \in \bbbc_\pm $ of $\lambda  $. Below we
analyze the structure of these degeneracies.

\subsection{The fundamental analytic solutions}\label{ssec:FAS}

The next step is to construct the fundamental analytic solutions
of (\ref{eq:4.1}). In our case this is done simply by combining
the blocks of Jost solutions with the same analytic properties:
\begin{eqnarray}\label{eq:6.3}
\chi ^+(x,\lambda ) &\equiv & \left(|\phi ^+\rangle , |\psi
^+\rangle\right)(x,\lambda) = \phi (x,\lambda ) \S^+(\lambda ) =
\psi (x,\lambda ) \T^-(\lambda ), \nonumber\\
\chi ^-(x,\lambda ) &\equiv & \left(|\psi ^-\rangle , |\phi
^-\rangle \right)(x,\lambda ) = \phi (x,\lambda ) \S^-(\lambda ) =
\psi (x,\lambda ) \T^+(\lambda ),
\end{eqnarray}
where the block-triangular functions $\S^\pm(\lambda ) $ and
$\T^\pm(\lambda ) $ are given by:
\begin{eqnarray}\label{eq:6.4}
\S^+(\lambda ) &=& \left( \begin{array}{cc} \openone  &
\d^-(\lambda )
\\ 0 & \c^+(\lambda ) \end{array}\right),\qquad
\T^-(\lambda ) =  \left( \begin{array}{cc} \a^+(\lambda ) & 0 \\
\b^+(\lambda ) & \openone  \end{array}\right), \nonumber\\
\S^-(\lambda ) &=&  \left( \begin{array}{cc} \c^-(\lambda ) & 0 \\
-\d^+(\lambda ) & \openone  \end{array}\right), \qquad
\T^+(\lambda ) = \left( \begin{array}{cc} \openone  &
-\b^-(\lambda )
\\ 0 & \a^-(\lambda ) \end{array}\right),
\end{eqnarray}
These triangular factors can be viewed also as generalized Gauss
decompositions (see \cite{Helg}) of $T(\lambda ) $ and its inverse:
\begin{eqnarray}\label{eq:7.1}
T(\lambda ) &=& \T^-(\lambda )\hat{\S}^+(\lambda ) =
\T^+(\lambda )\hat{\S}^-(\lambda ), \nonumber\\
\hat{T}(\lambda ) &=& \S^+(\lambda )\hat{\T}^-(\lambda ) =
\S^-(\lambda )\hat{\T}^+(\lambda ).
\end{eqnarray}
The relations between $\c^\pm(\lambda ) $, $\d^\pm(\lambda ) $ and
$\a^\pm(\lambda ) $, $\b^\pm(\lambda ) $ in eq. (\ref{eq:6.2})
ensure that equations (\ref{eq:7.1}) become identities. From eqs.
(\ref{eq:6.3}), (\ref{eq:6.4}) we derive:
\begin{eqnarray}\label{eq:9.4}
\chi ^+(x,\lambda ) &=& \chi ^-(x,\lambda ) G_0(\lambda ), \qquad
G_0(\lambda ) = \hat{D}^-(\lambda )(\openone +K^-(\lambda )),  \\
\label{eq:11.3} \chi ^-(x,\lambda ) &=& \chi ^+(x,\lambda )
\hat{G}_0(\lambda ), \qquad \hat{G}_0(\lambda ) =
\hat{D}^+(\lambda )(\openone -K^+(\lambda )),
\end{eqnarray}
valid for $\lambda \in \bbbr$, where
\begin{eqnarray}\label{eq:K-pm}
D^-(\lambda ) &=& \left( \begin{array}{cc} \c^-(\lambda ) &0 \\ 0
& \a^-(\lambda ) \end{array}\right), \qquad
K^-(\lambda ) = \left( \begin{array}{cc} 0& \d^-(\lambda ) \\
\b^+(\lambda ) & 0\end{array}\right), \\
D^+(\lambda ) &=& \left( \begin{array}{cc} \a^+(\lambda ) &0 \\ 0
& \c^+(\lambda ) \end{array}\right), \qquad
K^+(\lambda ) = \left( \begin{array}{cc} 0& \b^-(\lambda ) \\
\d^+(\lambda ) & 0\end{array}\right),
\end{eqnarray}
Obviously the block-diagonal factors $D^+(\lambda ) $ and
$D^-(\lambda )$ are matrix-valued analytic functions for $\lambda
\in \bbbc_\pm $. Another well known fact about the FAS $\chi ^\pm
(x,\lambda ) $ concerns their asymptotic behavior for $\lambda
\to\pm\infty  $, namely:
\begin{equation}\label{eq:12.1}
\lim_{\lambda \to\infty } X^\pm(x,\lambda ) =\openone ,
\end{equation}
where we have introduced:
\begin{equation}\label{eq:12.3}
X^\pm(x,\lambda ) = \chi ^\pm(x,\lambda ) \re^{\ri\lambda Jx}.
\end{equation}
In the derivations that follow the analyticity properties of
$X^\pm(x,\lambda ) $ for $\lambda \in\bbbc_\pm $ and eq.
(\ref{eq:12.1}) will play crucial role.

\subsection{The Gel'fand-Levitan-Marchenko equation}\label{ssec:GLM}

The first method to solve the ISM is based on the
Gel'fand-Levitan-Marchenko  (GLM) equation.  For the
Zakharov-Shabat system it is well known; for the block-matrix case
- see the pedagogical exposition in \cite{APT}. The GLM equation
is an integral equation for the transformation operator
$\mathcal{K} $ which relate the Jost solution $\psi (x,\lambda ) $
and the `plane waves' $\exp(-\ri\lambda Jx) $ as follows:

\begin{equation}\label{eq:TrOp}
\psi (x,\lambda ) \re^{\ri J\lambda x} = \openone +
\int_{x}^{\infty } \rd s \, \mathcal{K}(x,s) \re^{\ri\lambda
J(x-s)}.
\end{equation}

Then the GLM equation is a Volterra-type integral equation of the
form:
\begin{equation}\label{eq:GLM}
\mathcal{K}(x,y) = \mathcal{F}(x+y) + \int_{x}^{\infty } \rd s\,
\mathcal{K}(x,s) \mathcal{F}(s+y) ,
\end{equation}
where the kernel $\mathcal{F}(x) $ is expressed in terms of the
scattering data of the operator $L $ as follows:
\begin{eqnarray}\label{eq:F}
\mathcal{F}(x) &=& {1 \over 2\pi } \int_{-\infty }^{\infty }
\rd\lambda \left( \begin{array}{cc} 0 & \rho ^-(\lambda )
\exp(-\ri\lambda x) \\ \rho ^+(\lambda )\exp(\ri\lambda x) & 0
\end{array} \right) \nonumber \\ &+& \ri \sum_{j=1}^{N}
\left( \begin{array}{cc} 0 & \rho_j ^- \exp(-\ri\lambda_j^- x) \\
\rho_j^+\exp(\ri\lambda_j^+ x) & 0 \end{array} \right),
\end{eqnarray}
where $\rho ^\pm(\lambda ) $ are the reflection coefficients
introduced in (\ref{eq:rho-tau}) above, $\lambda _j^\pm $ are the
discrete eigenvalues of $L $ and the constants $\rho_j^\pm $
characterize the norming of the Jost solutions $\psi
^\pm(x,\lambda_j^\pm ) $, see \cite{APT}.

Let us now explain how the GLM eq. allows one to solve the ISP for
$L $. Indeed, given the scattering matrix $T(\lambda ) $ we can
construct the reflection coefficients $\rho ^\pm(\lambda ) $, the
coefficients $\rho _j^\pm $ and the discrete eigenvalues $\lambda
_j^\pm $, $j=1,\dots, N $. This provides us with the kernel
$\mathcal{F}(x) $ of the GLM eq. Since this is a Volterra type
equation it always has a solution. If one is able to solve it and
construct the transformation operator  $\mathcal{K}(x,y) $ then
the corresponding potential $Q(x)$  can be recovered through:
\begin{equation}\label{eq:Q}
Q(x) = \ri J[J,\mathcal{K}(x,x)].
\end{equation}
Thus, given the scattering data we recover the corresponding
potential $Q(x) $.

It is well known that in the reflectionless case $\rho ^+=\rho
^-=0 $ the kernel $\mathcal{F}(x) $ becomes degenerated and the
GLM equation can be solved exactly thus providing the
reflectionless potentials for $L $, which in turn are directly
related to the soliton solutions of the corresponding NLEE. We
will derive them below using the method, known now as the dressing
method \cite{Ma1,ZaSha,Za*Mi}.

\subsection{Reductions of $L $.}\label{ssec:2.2}

Typically one goes from the system (\ref{eq:3.1}) to the MNLS
equations (\ref{eq:3.2}) with $\epsilon =\pm 1 $ by imposing the
condition $\r = \epsilon \q^\dag $ which is known as reduction
condition. The theory for constructing such reductions was
proposed by A. Mikhailov in \cite{Mikh} and developed further in
\cite{vgn,vgrn,LoMi1}.

Here we analyze just $\bbbz_2 $-reduction, which are most widely
used in the literature:
\begin{equation}\label{eq:25.1}
B (U^\dag (x,t,\lambda^* ))B^{-1} = U(x,t,\lambda ), \qquad
U(x,t,\lambda ) = Q(x,t) -\lambda J, \qquad B^2=\openone .
\end{equation}
Choosing $B $ to be constant block-diagonal matrix:
\begin{equation}\label{eq:B}
B= \left( \begin{array}{cc} B_+ & 0 \\ 0 & B_- \end{array}\right),
\end{equation}
we find that eq. (\ref{eq:25.1}) leads to:
\begin{equation}\label{eq:25.2}
\r = B_- \q^\dag B_+^{-1}, \qquad \q = B_+ \r^\dag B_-^{-1},
\end{equation}
More precisely the reduction (\ref{eq:25.1}) means that:
\begin{equation}\label{eq:25.3}
(\chi ^+(x,t,\lambda ^*))^\dag = B^{-1}(\chi
^{\prime,-}(x,t,\lambda))^{-1} B,
\end{equation}
where $\chi ^+ $ and $\chi ^{\prime,-} $ are conveniently chosen
solutions of eq. (\ref{eq:4.1}). If we identify $\chi ^\pm $ as
the FAS (\ref{eq:6.3}) analytic for $\lambda \in \bbbc_\pm $ then
\begin{equation}\label{eq:25.3'}
(\chi ^+(x,t,\lambda ^*))^\dag = B^{-1}(\chi^{-}(x,t,\lambda)
\hat{D}^-(\lambda ) )^{-1} B.
\end{equation}
This result is derived easily by comparing the asymptotics of both
sides of eq. (\ref{eq:25.3'}) for $x\to\pm\infty  $. Skipping the
detail we will list here the basic consequences of the reductions
for the scattering data of $L $. For the scattering matrix we get:
\begin{equation}\label{eq:25.4}
T^\dag (t,\lambda ^*) = B\hat{T}(t,\lambda ) B^{-1},
\end{equation}
or in `components' (skipping the $t $-dependence):
\begin{eqnarray}\label{eq:25.5}
(\a^\pm (\lambda ^*))^\dag &=& B_\pm \c^\mp(\lambda ) \hat{B}_\pm,
\qquad (\b^\pm (\lambda ^*))^\dag = B_\pm \d^\mp(\lambda )
\hat{B}_\mp, \qquad
\lambda \in \bbbr,\nonumber\\
(\rho ^- (\lambda ))^\dag &=& B_- \rho ^+(\lambda ) \hat{B}_+,
\qquad (\tau^- (\lambda ))^\dag = B_+ \tau^+(\lambda ) \hat{B}_-, \\
(D^+(\lambda ^*))^\dag &=& B\hat{D}^-(\lambda )\hat{B}.\nonumber
\end{eqnarray}
For other interesting choices leading to new reductions of $N
$-wave and MNLS-type equations, see
\cite{vgn,vgrn,varna04,manev04}.

\subsection{The Riemann-Hilbert problem}\label{ssec:RHP}

The eqs. (\ref{eq:9.4}) and (\ref{eq:11.3}) can be written down
as:
\begin{equation}\label{eq:12.4}
X^+(x,\lambda ) = X^-(x,\lambda ) \hat{D}^-(\lambda ) (\openone
+K^-(x,\lambda )), \qquad \lambda \in \bbbr
\end{equation}
or
\begin{equation}\label{eq:12.5}
X^-(x,\lambda ) = X^+(x,\lambda ) \hat{D}^+(\lambda ) (\openone
-K^+(x,\lambda )), \qquad \lambda \in \bbbr
\end{equation}
where
\begin{equation}\label{eq:12.6}
K^\pm(x,\lambda ) = \re^{-\ri\lambda Jx} K^\pm(\lambda
)\re^{\ri\lambda Jx}.
\end{equation}

Eq. (\ref{eq:12.4}) (resp. eq. (\ref{eq:12.5})) combined with
(\ref{eq:12.1}) is known in the literature \cite{Gakhov} as a
Riemann-Hilbert problem (RHP) with canonical normalization. It is
well known that RHP with canonical normalization has unique
regular solution; the matrix-valued solutions $X_0^+(x,\lambda ) $
and $X_0^-(x,\lambda ) $ of (\ref{eq:12.4}), (\ref{eq:12.1}) is
called regular if $\det X_0^\pm(x,\lambda ) $ does not vanish for
any $\lambda \in\bbbc_\pm $.

\begin{figure}[tb]
\includegraphics*{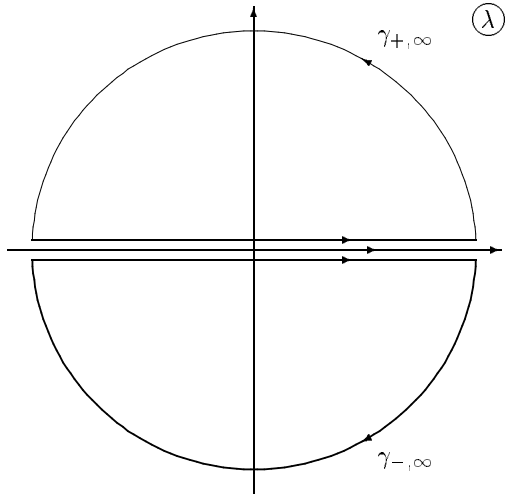}
\caption{The contours $\gamma _\pm =\bbbr\cup\gamma_{\pm\infty
}$.} \label{fig:1}

\end{figure}

Let us now apply the contour-integration method to derive the
integral decompositions of $X^\pm(x,\lambda ) $. To this end we
consider the contour integral:
\begin{equation}\label{eq:J_1}
\mathcal{J}_1(\lambda ) = {1 \over 2\pi \ri} \oint_{\gamma _+}
{\rd\mu \over \mu -\lambda } X^+(x,\mu ) - {1 \over 2\pi \ri}
\oint_{\gamma _-} {\rd\mu \over \mu -\lambda } X^-(x,\mu
)\hat{D}^-(\mu ),
\end{equation}
where $\lambda \in\bbbc_+ $ and the contours $\gamma _\pm $ are
shown on fig.~\ref{fig:1}.

First we evaluate $\mathcal{J}_1(\lambda ) $ by Cauchy theorem.
The result is:
\begin{equation}\label{eq:J_11}
\mathcal{J}_1(\lambda ) = X^+(x,\lambda ) + \sum_{j =1}^{N}
\Res_{\mu =\lambda _j^-} {X^-(x,\mu )\hat{D}^-(\mu ) \over \mu
-\lambda  }.
\end{equation}
We can also evaluate $\mathcal{J}_1(\lambda ) $ by integrating
along the contours. In integrating along the infinite semi-circles
of $\gamma _{\pm,\infty } $ we use the asymptotic behavior of
$X^\pm(x,\lambda ) $ for $\lambda \to\infty $. The result is:
\begin{equation}\label{eq:J_12}
\mathcal{J}_1(\lambda ) = \openone + {1 \over 2\pi \ri}
\int_{-\infty} ^{\infty }{\rd\mu \over \mu -\lambda } X^-(x,\mu )
\hat{D}^-(\mu ) K^-(\mu),
\end{equation}
where in evaluating the integrand  we made use of eq.
(\ref{eq:12.4}). Equating the right hand sides of  (\ref{eq:J_11})
and (\ref{eq:J_12}) we get the following integral decomposition
for $X^+(x,\lambda ) $:
\begin{eqnarray}\label{eq:X^+}
X^+(x,\lambda ) &=& \openone - \sum_{j =1}^{N} \Res_{\mu=\lambda
_j^-}
{X^-(x,\mu )\hat{D}^-(\mu ) \over \mu -\lambda  } \nonumber\\
&+&{1 \over 2\pi \ri} \int_{-\infty }^{\infty } {\rd\mu\over \mu
-\lambda } X^-(x,\mu ) \hat{D}^-(\mu ) K^-(\mu ),
\end{eqnarray}

Quite analogously we derive the decomposition for $X^-(x,\lambda )
$:
\begin{eqnarray}\label{eq:X^-}
X^-(x,\lambda ) &=& \openone - \sum_{j =1}^{N} \Res_{\mu =\lambda
_j^+} {X^+(x,\mu )\hat{D}^+(\mu ) \over \mu -\lambda  }
\nonumber\\
&+&{1 \over 2\pi \ri} \int_{-\infty }^{\infty } {\rd\mu\over \mu
-\lambda } X^+(x,\mu ) \hat{D}^+(\mu ) K^+(\mu ),
\end{eqnarray}

Equations (\ref{eq:X^+}), (\ref{eq:X^-}) can be viewed as a set of
singular integral equations which are equivalent to the RHP. For
the MNLS these were first derived in \cite{Ma1}.

\subsection{The minimal set of scattering data}\label{ssec:2x}

Obviously, given the potential $Q(x) $ one can solve the integral
equations for the Jost solutions which determine them uniquely.
The Jost solutions in turn determine uniquely the scattering
matrix $T(\lambda ) $ and its inverse $\hat{T}(\lambda ) $. But
$Q(x) $ contains at most $2mn $ independent complex-valued
functions of $x $. Thus it is natural to expect that at most $2mn
$ of the coefficients in $T(\lambda ) $ for $\lambda \in \bbbr$
will be independent; the rest must be functions of those.

The set of independent coefficients of $T(\lambda ) $ are known as
the minimal set of scattering data. As such we may use any of the
following two sets $\mathcal{T}_{i} \equiv \mathcal{T}_{i,\rm
c}\cup \mathcal{T}_{i,\rm d} $:
\begin{eqnarray}\label{eq:T_12}
\mathcal{T}_{1,\rm c} \equiv \left\{ \rho ^+(\lambda ), \rho
^-(\lambda ), \quad \lambda \in \bbbr\right\},    \qquad
\mathcal{T}_{1,\rm d} \equiv \left\{\rho_j^\pm, \lambda _j^\pm
\right\}_{j=1}^{N}, \nonumber\\
\mathcal{T}_{2,\rm c} \equiv \left\{ \tau^+(\lambda ),
\tau^-(\lambda ), \quad \lambda \in \bbbr\right\},    \qquad
\mathcal{T}_{1,\rm d} \equiv \left\{\tau_j^\pm, \lambda _j^\pm
\right\}_{j=1}^{N},
\end{eqnarray}
where the reflection coefficients $\rho ^\pm(\lambda ) $ and
$\tau^\pm(\lambda ) $ were introduced in eq. (\ref{eq:6.2}),
$\lambda _j^\pm $ are (simple) discrete eigenvalues of $L $ and
$\rho_j^\pm $ and $\tau_j^\pm $ characterize the norming constants
of the corresponding Jost  solutions.

The reflection coefficients $\rho ^\pm(\lambda ) $ and
$\tau^\pm(\lambda ) $ are defined only on the real $\lambda
$-axis, while the diagonal blocks $\a^\pm(\lambda ) $ and
$\c^\pm(\lambda ) $ (or, equivalently, $D^\pm(\lambda ) $) allow
analytic extensions for $\lambda \in \bbbc_\pm $. From the
equations (\ref{eq:6.2}) there follows that:
\begin{eqnarray}\label{eq:2x2-1}
\a^+(\lambda )\c^-(\lambda ) = (\openone +\rho ^-\rho ^+(\lambda
))^{-1}, \qquad \a^-(\lambda )\c^+(\lambda) =(\openone +\rho^+\rho
^-(\lambda
))^{-1},\\
\label{eq:2x2-2} \c^-(\lambda )\a^+(\lambda ) = (\openone
+\tau^+\tau ^-(\lambda ))^{-1}, \qquad \c^+(\lambda )\a^-(\lambda
) = (\openone +\tau ^-\tau ^+(\lambda ))^{-1},
\end{eqnarray}

Given $\mathcal{T}_1 $ (resp., $\mathcal{T}_2 $) we determine the
right hand sides of  (\ref{eq:2x2-1}) (resp. (\ref{eq:2x2-2})) for
$\lambda \in \bbbr $. Combined with the facts about the limits:
\begin{eqnarray}\label{eq:2x2-3}
\lim_{\lambda \to\infty } \a^+(\lambda ) =\openone , \qquad
\lim_{\lambda \to\infty } \c^-(\lambda ) =\openone ,
\nonumber\\
\lim_{\lambda \to\infty } \a^-(\lambda ) =\openone , \qquad
\lim_{\lambda \to\infty } \c^+(\lambda ) =\openone ,
\end{eqnarray}
each of the relations (\ref{eq:2x2-1}), (\ref{eq:2x2-2}) can be
viewed as a RHP with canonical normalization. Such RHP can be
solved explicitly in the one-component case (provided we know the
locations of their zeroes) by using the Plemelj-Sokhotsky formulae
\cite{Gakhov}.
These zeroes are in fact the discrete eigenvalues of $L
$. One possibility to make use of these facts is to take log of
the determinants of both sides of (\ref{eq:2x2-1}) getting:
\begin{equation}\label{eq:2x2-5}
A^+(\lambda )+C^-(\lambda ) = - \ln \det (\openone +\rho ^-\rho
^+(\lambda ), \qquad \lambda \in \bbbr,
\end{equation}
where
\begin{equation}\label{eq:2x2-5'}
A^\pm (\lambda ) = \ln \det \a^\pm(\lambda ), \qquad C^\pm
(\lambda ) = \ln \det \c^\pm(\lambda ).
\end{equation}
Then Plemelj-Sokhotsky formulae  allows us to recover
$A^\pm(\lambda ) $ and $C^\pm(\lambda ) $:
\begin{eqnarray}\label{eq:2x2-6}
\mathcal{A}(\lambda ) = {\ri \over 2\pi } \int_{-\infty }^{\infty
} { \rd\mu\over \mu -\lambda } \ln \det (\openone +\rho ^-\rho
^+(\mu )) + \sum_{j=1}^{N} \ln {\lambda -\lambda _j^+ \over
\lambda -\lambda _j^- },
\end{eqnarray}
where $\mathcal{A}(\lambda )=A^+(\lambda ) $ for $\lambda \in
\bbbc_+ $ and $\mathcal{A}(\lambda )=-C^-(\lambda ) $ for $\lambda
\in \bbbc_-$. In deriving (\ref{eq:2x2-6}) we have also assumed
that $\lambda _j^\pm $ are simple zeroes of $A^\pm(\lambda ) $ and
$C^\pm(\lambda ) $.

\begin{remark}\label{rem:2'}
The sets (\ref{eq:T_12}) were first derived for $m=n=1 $ in
\cite{DJK1}, see also \cite{GKh1}. Here we are using the gauge
covariant formulation developed in \cite{vg-ya}.
\end{remark}

If we impose the reduction condition (\ref{eq:25.5}) then we get
the following constraints on the sets $\mathcal{T}_{1,2} $:
\begin{eqnarray}\label{eq:rho-1}
\rho ^-(\lambda )=(B_-\rho ^+(\lambda )B_+)^\dag, \qquad \rho_j
^-=(B_-\rho_j ^+B_+)^\dag , \qquad \lambda _j^-=(\lambda _j^+)^*,
\\ \label{eq:tau-1} \tau^-(\lambda )=(B_+\tau ^+(\lambda )B_-)^\dag,
\qquad \tau_j ^-=(B_+\tau_j ^+B_-)^\dag , \qquad \lambda_j^-
=(\lambda _j^+)^*,
\end{eqnarray}
where $ j=1,\dots, N, $

\begin{remark}\label{rem:3}
For certain choices of the reduction conditions (\ref{eq:Q}), such
as $Q=-Q^\dag $ the generalized Zakharov-Shabat system $L(\lambda
)\psi =0 $ can be written down as an eigenvalue problem
$\mathcal{L}\psi =\lambda \psi (x,\lambda ) $ where $\mathcal{L} $
is a self-adjoint operator. The continuous spectrum of
$\mathcal{L} $ fills up the whole real $\lambda $-axis thus
`leaving no space' for discrete eigenvalues. Such Lax operators
have no discrete spectrum and the corresponding MNLS do not have
soliton solutions.

\end{remark}

The full RHP's (\ref{eq:2x2-1}), (\ref{eq:2x2-2}) for $n $ and
$m>1 $ do not allow explicit solutions.  However, from the general
theory of RHP \cite{Gakhov} one may conclude that
(\ref{eq:2x2-1}), (\ref{eq:2x2-2}) allow unique solutions provided
the number and types of the zeroes $\lambda _j^\pm $ are properly
chosen.

Thus we can outline a procedure which allows one to reconstruct
not only $T(\lambda ) $ and $\hat{T}(\lambda ) $ and the
corresponding potential $Q(x) $ from each of the sets
$\mathcal{T}_i $, $i=1,2 $:

\begin{description}

\item [i)] Given $\mathcal{T}_1 $ (resp. $\mathcal{T}_2 $) solve
the RHP (\ref{eq:2x2-1}) (resp. (\ref{eq:2x2-2})) and construct
$\a^\pm(\lambda ) $ and $\c^\pm(\lambda ) $ for $\lambda
\in\bbbc_\pm $.

\item [ii)] Given $\mathcal{T}_1 $ we determine $\b^\pm(\lambda )
$ and $\d^\pm(\lambda ) $ as:
\begin{equation}\label{eq:2x2-7}
\b^\pm(\lambda ) =\rho ^\pm(\lambda )\a^\pm(\lambda ) , \qquad
\d^\pm(\lambda ) =\c^\pm(\lambda )\rho ^\pm(\lambda ) ,
\end{equation}
or if $\mathcal{T}_2 $ is known then
\begin{equation}\label{eq:2x2-7'}
\b^\pm(\lambda ) =\a^\pm(\lambda )\tau ^\pm(\lambda ) ,\qquad
\d^\pm(\lambda ) =\tau ^\pm(\lambda )\c^\pm(\lambda ) .
\end{equation}

\item [iii)] The potential $Q(x) $ can be recovered from
$\mathcal{T}_1 $ by solving the GLM eq. and using eq.
(\ref{eq:F}).

\end{description}

Another method for reconstructing $Q(x) $ from $\mathcal{T}_j $
uses the interpretation of the ISM as generalized Fourier
transform, see \cite{AKNS,KN79,DJK1,GKh1,G,GYa,vg-ya} and Section
4 below.

\section{The RHP and the inverse scattering
problem}\label{ssec:RHP-ISP}

There are two pioneer results that enhanced the development of the
ISM. The first is the existence and explicit construction of the
FAS for rather general class of Lax operators
\cite{Sh,Sha,ZaMa,BeCo,BeSat,GYa}.  The second is the discovery of
the equivalence of the ISP for $L $ to a RHP for the FAS
\cite{Sha,ZaSha,ZaMik,Za*Mi}.

Here we are considering rather special Lax operators for which the
construction of the FAS does not require special efforts. Here we
first prove the equivalence between RHP and the spectral problem
$L $. Next we show how the FAS can be used to construct the
resolvent of $L $ and for the analysis of its spectral properties.

The use of the RHP rather than the GLM equations allowed for
important effective extensions of the ISM; it allowed the
treatment of generic Lax operators with rational dependence on
$\lambda  $ \cite{ZaMik,Za*Mi,LoMi1}, as well as of ones related
to simple Lie algebras \cite{G,vgn,vgrn,varna04,G-cm,GYa}.  In the
next subsections we will make use of these developments to
describe two types of soliton solutions for the MNLS.

\subsection{Equivalence of RHP to ISP}\label{ssec:3.1}

The next important step is the possibility to reduce the solution
of the ISP for the generalized Zakharov-Shabat system to a (local) RHP.
Indeed the relation (\ref{eq:9.4}) can be rewritten as:
\begin{subequations}\label{eq:23.1}
\begin{eqnarray}\label{eq:23.1a}
X^+(x,t,\lambda )&=& X^-(x,t,\lambda ) G(x,t,\lambda ), \qquad
\lambda \in \bbbr, \\
\label{eq:23.1b} G(x,t,\lambda )&=& \re^{-\ri(\lambda Jx-f(\lambda
)t)} G_0(\lambda )
\re^{\ri(\lambda Jx-f(\lambda )t)}, \\
\label{eq:23.1c} G_0(\lambda ) &=& \left.
\hat{S}^-(\lambda,t)S^+(\lambda,t) \right|_{t=0};
\end{eqnarray}
\end{subequations}
in other words the sewing function $G(x,t,\lambda ) $ satisfies
the equations:
\begin{eqnarray}\label{eq:23.2}
i {\rd G\over \rd x } -\lambda [J,G(x,t,\lambda )]=0, \qquad i
{\rd G\over \rd t } +[f(\lambda) ,G(x,t,\lambda )]=0,
\end{eqnarray}
Here $f(\lambda) \in \mathfrak{h}$ determines the dispersion law
of the NLEE.

\begin{theorem}[\cite{Sha}]\label{th:Sha} Let $X^+(x,t,\lambda ) $ and
$X^-(x,t,\lambda )$ be solutions to the RHP (\ref{eq:23.1}) with
canonical normalization (\ref{eq:12.1}) allowing analytic
extension in $\lambda  $ for $\lambda \in \bbbc_\pm $
respectively. Then $\chi^\pm(x,t,\lambda ) =X^\pm(x,t,\lambda
)\re^{\ri\lambda Jx} $ are fundamental analytic solutions of both
operators $L $ and $M $, i.e. satisfy  eqs.  (\ref{eq:4.1}),
(\ref{eq:4.1M}) with
\begin{equation}\label{eq:23.3}
Q(x,t) = \lim_{\lambda  \to\infty } \lambda \left(J -
X^\pm(x,t,\lambda ) J \hat{X}^\pm(x,t,\lambda )\right).
\end{equation}
\end{theorem}

\begin{proof}
Let us assume that $X^\pm(x,t,\lambda ) $ are regular solutions to
the RHP and let us introduce the function:
\begin{equation}\label{eq:24.1}
g^\pm(x,t,\lambda ) = \ri { \rd X^\pm \over \rd x }
\hat{X}^\pm(x,t,\lambda) + \lambda X^\pm(x,t,\lambda )J
\hat{X}^\pm(x,t,\lambda ).
\end{equation}
If $X^\pm(x,t,\lambda )$ are regular then neither
$X^\pm(x,t,\lambda )$ nor their inverse $\hat{X}^\pm(x,t,\lambda )
$ have singularities in their regions of analyticity $\lambda \in
\bbbc_\pm $. Then the functions $g^\pm(x,t,\lambda ) $ also will
be regular for all $\lambda \in \bbbc_\pm$. Besides:
\begin{equation}\label{eq:24.2}
\lim_{\lambda \to\infty }g^+(x,t,\lambda ) = \lim_{\lambda
\to\infty } g^-(x,t,\lambda ) = \lambda J.
\end{equation}
The crucial step in the proof \cite{ZaSha} is based on the
chain of relations:
\begin{eqnarray}\label{eq:24.3}
&& g^+(x,t,\lambda ) \mathop{=}\limits^{(\ref{eq:23.1})} \ri {\rd
(X^-G ) \over \rd x } \hat{G}\hat{X}^-(x,t,\lambda ) + \lambda
X^-G
J\hat{G} \hat{X}^-(x,t,\lambda ) \nonumber\\
&&\qquad = \ri {\rd X^- \over \rd x }\hat{X}^-(x,t,\lambda ) + X^-
\left( \ri {\rd G \over \rd x }\hat{G} + \lambda GJ\hat{G}\right)
\hat{X}^-(x,t,\lambda ) \nonumber\\
&& \qquad \mathop{=}\limits^{(\ref{eq:23.2})} \ri {\rd X^- \over
\rd x }\hat{X}^-(x,t,\lambda ) + X^- \left( \lambda [J,G]\hat{G} +
\lambda GJ\hat{G}\right) \hat{X}^-(x,t,\lambda )
\nonumber\\
&& \qquad = \ri {\rd X^- \over \rd x }\hat{X}^-(x,t,\lambda ) +
\lambda
X^- J\hat{X}^-(x,t,\lambda ) \nonumber\\
&&  \qquad \equiv g^-(x,t,\lambda ), \qquad \lambda \in \bbbr.
\end{eqnarray}
Thus we conclude that $g^+(x,t,\lambda )=g^-(x,t,\lambda ) $ is a
function analytic in the whole complex $\lambda  $-plane except in
the neighborhood of $\lambda \to\infty  $ where $g^+(x,t,\lambda )
$ tends to $\lambda J $, (\ref{eq:24.2}). Next from Liouville
theorem we conclude that the difference $g^+(x,t,\lambda )
-\lambda J$ is a constant with respect to $\lambda  $; if we
denote this `constant' by $-Q(x,t) $ we get:
\begin{equation}\label{eq:24.4}
g^+(x,t,\lambda ) -\lambda J = -Q(x,t).
\end{equation}
Using the definition of $g^+(x,t,\lambda ) $ (\ref{eq:24.1}) we
find that $X^\pm(x,t,\lambda ) $ satisfy
\begin{equation}\label{eq:Dr2.3}
\ri {\rd X^\pm \over \rd x } +Q(x,t)X^\pm(x,t,\lambda) - \lambda
[J,X^\pm(x,t,\lambda)]=0,
\end{equation}
i.e. that $\chi^\pm(x,t,\lambda)$ is a fundamental solution to
$L$. The relation between $Q(x,t) $ and $X^\pm(x,t,\lambda ) $
(\ref{eq:23.3}) is obtained by taking the limit of the left-hand
sides of (\ref{eq:24.4}) for $\lambda \to\infty  $ and using the
asymptotic expansions of $X^\pm(x,\lambda)$ and
$\hat{X}^\pm(x,\lambda)$:
\begin{eqnarray}\label{eq:Dr2.4}
X^\pm(x,t,\lambda) = \openone + \sum_{s=1}^\infty \lambda^{-s}
X_s^\pm(x,t), \nonumber \\
\hat{X}^\pm(x,t,\lambda) = \openone + \sum_{s=1}^\infty
\lambda^{-s} \hat{X}_s^\pm(x,t),
\end{eqnarray}
we get:
\begin{equation}\label{eq:DR2.5}
Q(x,t) = \lim_{\lambda\to\infty} \lambda (J - X^\pm J\hat{X}^\pm
(x,t,\lambda)] = [J, X_1^\pm(x,t)].
\end{equation}
Here we also used the fact that $ X_1^\pm(x,t)=-
\hat{X}_1^\pm(x,t)$, see eq. (\ref{eq:Dr2.4}).

Arguments along the same line applied to the functions
$h^\pm(x,t,\lambda )$
\begin{equation}\label{eq:24.5}
h^\pm(x,t,\lambda )= \ri {\rd X^\pm\over \rd t }
\hat{X}^\pm(x,t,\lambda) +2\lambda^2 X^\pm(x,t,\lambda )J
\hat{X}^\pm(x,t,\lambda ),
\end{equation}
can be used to prove that $\chi ^\pm(x,t,\lambda ) $ are
fundamental solutions also of the operator $M $ (\ref{eq:4.1M}).
Indeed, repeating the above arguments for $h^\pm(x,t,\lambda ) $
we find that $h^+(x,t,\lambda )=h^-(x,t,\lambda ) $ is a function
analytic everywhere in $\bbbc $ except at $\lambda \to\infty $
where it behaves like $2\lambda ^2J$. From Liouville theorem it
follows that $h^\pm(x,t,\lambda )-2\lambda ^2J $ is a linear
function of $\lambda $ equal to $-V_0(x,t) -\lambda V_1(x,t) $.
Thus:
\begin{equation}\label{eq:24.6}
\ri {\rd X^\pm  \over \rd t } + (V_0(x,t) +\lambda V_1(x,t))
X^\pm(x,t,\lambda ) - [2\lambda^2J, X^\pm(x,t,\lambda )]=0,
\end{equation}

To conclude the proof we have to account for possible zeroes and
pole singularities  of $X^\pm(x,t,\lambda ) $ at the points
$\lambda _j^\pm $.  Below we derive the structure of these
singularities and show that they do not influence the functions
$g^\pm(x,t,\lambda ) $ and $h^\pm(x,t,\lambda ) $.
\end{proof}

\subsection{The dressing Zakharov-Shabat method for the symmetric
spaces
 $SU(n+m)/S(U(n)\otimes U(m))$.}\label{ssec:DrZS}

Let us outline how one can, starting from a given regular
solutions $X_0^\pm(x,t,\lambda ) $ of the RHP, construct new
singular solutions $X^{\pm}(x,t,\lambda ) $ having zeroes
(singularities) at the prescribed points $\lambda _j^\pm \in
\bbbc_\pm $.  The structure of these singularities are determined
by the dressing factor $u_j(x,t,\lambda ) $:
\begin{equation}\label{eq:26.1}
\xi^{\pm}(x,t,\lambda ) = u_j(x,t,\lambda )
\xi_0^{\pm}(x,t,\lambda ) w_{j,\pm}^{-1}(\lambda ),
\end{equation}
which in our case has a simple fraction-linear dependence on
$\lambda$:
\begin{eqnarray}\label{eq:26.2}
u_j(x,t,\lambda ) &=& \openone + (c_j(\lambda ) -1)P_j(x,t),
\qquad c_j(\lambda ) = {\lambda  -\lambda _j^+ \over \lambda
-\lambda _j^- }.
\end{eqnarray}
Here $P_j(x,t) $ is a projector $P_j^2=P_j $,
$w_{j,+}(\lambda)=\openone$ and $w_{j,-}(\lambda ) $ will be
defined below.

It is well known that, if $Q(x,t)-\lambda J $ takes values in the
Lie algebra $\mathfrak{g} $ then $X^\pm(x,t,\lambda ) $ must take
values in the corresponding Lie group $\mathfrak{G} $.  Therefore
the dressing factor $u_j(x,t,\lambda ) $ must also be element of
the same group. But assuming $P_j(x,t) $ is a projector we can
derive that
\begin{equation}\label{eq:det-u}
\det u(x,t,\lambda ) = (c_j(\lambda ))^{r_j}, \qquad r_j=\rank
P_j(x,t).
\end{equation}
Obviously the ansatz (\ref{eq:26.2}) is compatible with
$\mathfrak{G}\simeq GL(n+m) $, $SL(n+m) $ and $U(n+m) $, $SU(n+m)
$. The last two possibilities are realized provided proper
reduction conditions like (\ref{eq:rho-1}), (\ref{eq:tau-1}) are
imposed on $\lambda _j^\pm $ and $P_j(x,t)=BP_j^\dag (x,t)B^{-1}
$.

Usually $P_j(x) $ is chosen to be of rank 1. We will consider
slightly more general case when $\rank P_j(x)=r_j\geq 1 $; $r_j $
is the multiplicity of the corresponding eigenvalues $\lambda
_j^\pm $ and $r_j\leq \min (n,m) $.  Then $P_j(x) $ can be written
in the form:
\begin{equation}\label{eq:26.3}
P_j(x) = |n_{j}\rangle \left( \langle m_{j}| n_{j}\rangle
\right)^{-1} \langle m_{j}| ,
\end{equation}
where the collections of $r_j $ bra- (resp. ket-) eigenvectors
$\langle m_{j}| $ (resp. $|n_{j}\rangle $) can be viewed also as
rectangular $(n+m)\times r_j $ (resp. $r_j\times (n+m)$) matrices.
Then $\langle m_{j}| n_{j}\rangle $ will be quadratic $r_j\times
r_j $ matrix which we assume to be nondegenerate.  If $r_j=1 $ eq.
(\ref{eq:26.3}) provides the standard expression for rank 1
projector. It is easy to check that the left hand side of
(\ref{eq:26.3}) satisfies identically the relation $P_j^2=P_j $.

{}From (\ref{eq:26.1}) there follows that the dressing factor
$u(x,t,\lambda )$ satisfies the equation:
\begin{equation}\label{eq:u-eq}
\ri {\rd u \over \rd x } + Q(x,t) u(x,t,\lambda ) - u(x,t,\lambda
) Q_0(x,t) - \lambda [J,u(x,t,\lambda )] =0,
\end{equation}
where $Q_0(x,t) $  (resp. $Q(x,t)$) is the potential related to
the regular $\chi_0^\pm(x,t,\lambda ) $ (resp. singular
$\chi^\pm(x,t,\lambda ) $) solution  of the RHP.

Next we insert the anzats (\ref{eq:26.2}) and request that it
holds identically with respect to $\lambda  $. To this end it is
enough that eq. (\ref{eq:26.2}) holds true for $\lambda =\lambda
_j^+ $, $\lambda \to \lambda _j^- $ and $\lambda \to \infty  $.
The first two conditions lead to the following equations for
$\langle m_{j}| $ and $|n_{j}\rangle $:
\begin{eqnarray}\label{eq:27.1}
\ri {\rd |n_{j}\rangle \over \rd x } + U^{(0)}(x,t,\lambda
_j^+)|n_{j} \rangle =0, &\quad &\ri {\rd |n_j\rangle \over \rd t }
+
V^{(0)}(x,t,\lambda_j^+) |n_{j}\rangle =0,\\
\ri {\rd\langle m_{j} |\over \rd x } - \langle m_{j}|
U^{(0)}(x,t,\lambda _j^-)=0, &\quad & \ri {\rd\langle m_{j} |\over
\rd t } -\langle m_{j}|
V^{(0)}(x,t,\lambda_j^-)=0,\\
U^{(0)}(x,t,\lambda )=Q_0(x,t) - \lambda J, &\quad &
V^{(0)}(x,t,\lambda )=\left. V(x,t,\lambda )\right|_{Q=Q_0}.
\end{eqnarray}
Here $V^{(0)}(x,t,\lambda ) $ is obtained from $V(x,t,\lambda ) $
(see (\ref{eq:4.1M}), (\ref{eq:4.1V})) replacing $Q(x,t) $ by
$Q_0(x,t)$. This construction is well defined also in the case
when $\chi _0^\pm(x,\lambda ) $ are singular solutions to the RHP,
provided they are regular for $\lambda =\lambda _j^\pm $. In order
to avoid technicalities in what follows we will treat only the
case of one pair of discrete eigenvalues. From eqs.
(\ref{eq:27.1}) there follows that:
\begin{equation}\label{eq:27.1'}
|n_{j}\rangle =\chi_{0j}^{+}(x,t) |n_{j}^0\rangle , \qquad \langle
m_{j}| = \langle m_{j}^0 |\hat{\chi}_{0j}^{-}(x,t), \qquad
\chi_{0j}^{\pm}(x,t)=\chi_0^{\pm}(x,t,\lambda_j^\pm ).
\end{equation}
Note that since $\chi_{0}^{\pm}(x,t)$ are solutions of the regular
RHP then $\chi_{0j}^{\pm}(x,t)$ exist and are nondegenerate. The
equation (\ref{eq:23.3}) considered for $\lambda \to\infty  $
gives the following relation between $Q_0(x,t) $, $Q(x,t) $ and
$P(x,t) $:
\begin{eqnarray}\label{eq:27.2}
Q(x,t) &=& Q_0(x,t) + \lim_{\lambda \to\infty } \lambda (J -
u_j(x,t,\lambda ) J\hat{u}_j(x,t,\lambda )) \nonumber\\
&=& Q_0(x,t) - (\lambda _j^+ - \lambda _j^-) [J,P_j(x,t)].
\end{eqnarray}

Thus starting from a given regular solution of the RHP (and
related solution $Q_0(x,t) $ to the NLEE) we can construct a
singular solution to the RHP and a new solution $Q(x,t) $ of the
NLEE depending on the $\lambda _j^\pm $ and on the eigenvectors of
$P_j(x) $. If we start from the trivial solution $Q_0(x,t)=0 $ of
the NLEE then we will get the one-soliton solution of the NLEE.
Repeating the procedure $N $ times we can get the $N $-soliton
solution of the NLEE.

With the explicit formulae for $P_j(x) $ and using (\ref{eq:26.1})
we can establish the relationship between the scattering data of
the regular RHP and the corresponding singular one. The dressing
factor $u_j(x,t,\lambda ) $ is determined by the collections of
constant polarization vectors
\begin{equation}\label{eq:n-m}
|n_{j}^0\rangle = \left( \begin{array}{c} |n_{0,j}^1\rangle \\
|n_{0,j}^2\rangle \end{array} \right), \qquad \langle m_{j}^0| =
\left( \langle m_{0,j}^1|, \langle m_{0,j}^2| \right),
\end{equation}
which can not be quite arbitrary. They should be such that the
constant $r_j\times r_j $ matrices $\langle
m_{0,j}^1|n_{0,j}^1\rangle $ and $\langle
m_{0,j}^2|n_{0,j}^2\rangle $ are nondegenerate.

Now we can evaluate the limits of the dressing factor $u(x,\lambda
) $ for $x\to\pm\infty  $ and then derive the interrelations
between the scattering matrices $T_0(\lambda ) $ and $T_1(\lambda
) $  corresponding to the potentials $Q_0(x,t) $ and $Q_1(x,t) $.
In what follows the elements of the scattering matrix $T_0(\lambda
) $ will be denoted by the same letters as the ones of $T(\lambda
) $ but with additional subscript $0 $.

Using the explicit formulae (\ref{eq:27.1'}), (\ref{eq:6.3}),
(\ref{eq:6.4}) we get:
\begin{eqnarray}\label{eq:P_pm}
P_j^\pm &=& \lim_{x\to\pm\infty } P_j(x); \qquad P_j^+ =
\left(\begin{array}{cc} P_{11,j}^{+} & 0 \\ 0 & 0
\end{array}\right), \qquad  P_j^-(x) = \left(\begin{array}{cc} 0 &
0 \\ 0 &
P_{22,j}^{-} \end{array}\right), \\
&& P_{11,j}^{+} = \a_0^+(\lambda _j^+)|n_{0,j}^{1} \rangle
\left(\langle m_{0,j}^{1}| \a_0^+(\lambda _j^+)| n_{0,j}^{1}
\rangle \right)^{-1}
\langle m_{0,j}^{1}|, \\
&&P_{22,j}^{-} = \c_0^+(\lambda_j^+)|n_{0,j}^{2} \rangle
\left(\langle m_{0,j}^{2}| \c_0^+(\lambda _j^+)|n_{0,j}^{2}
\rangle \right)^{-1} \langle m_{0,j}^{2}|,\nonumber
\end{eqnarray}
where obviously $\rank P_{11,j}^{-} = \rank P_{22,j}^{-} =r_j $.
Therefore the limits of the dressing factor are given by:
\begin{eqnarray}\label{eq:15.2}
\lim_{x\to\infty } u(x,\lambda ) &=& \left( \begin{array}{cc}
u_{11,j}^{+} & 0 \\ 0 & \openone \end{array}\right), \qquad
u_{11,j}^{+}= \openone  + (c_j(\lambda )-1)P_{11,j}^{+}, \\
\label{eq:15.4} \lim_{x\to -\infty } u(x,\lambda ) &=& \left(
\begin{array}{cc}
 \openone  & 0 \\ 0 & u_{22,j}^{-} \end{array}\right), \qquad
u_{22,j}^{-}= \openone  + (c_j(\lambda )-1)P_{22,j}^{-}.
\end{eqnarray}

Next we have to determine the matrices $w_{j,\pm}(\lambda ) $ in
the right hand side of eq. (\ref{eq:26.1}) so that the asymptotics
of the singular solution are compatible with (\ref{eq:6.3}),
(\ref{eq:6.4}). This holds true if
\begin{equation}\label{eq:15.11}
w_{j,+}(\lambda )=\openone  , \qquad w_{j,-}(\lambda )\equiv
W(\lambda ) =  \left(\begin{array}{cc} u_{11,j}^{+} & 0 \\ 0 &
u_{22,j}^{-}
\end{array}\right).
\end{equation}
In addition we find the following relations between the scattering
matrices:
\begin{eqnarray}\label{eq:16.2}
T(\lambda ) &=& u_{j}^{+} T_0(\lambda ) \hat{u}_{j}^{-},
\end{eqnarray}
or `in components':
\begin{subequations}\label{eq:16.2'}
\begin{eqnarray}\label{eq:16.2'a}
\a^+(\lambda ) &=& u_{11,j}^{+} \a_0^+(\lambda ), \qquad
\b^-(\lambda ) = u_{11,j}^{+} \b_0^-(\lambda )\hat{u}_{22,j}^{-}, \\
\label{eq:16.2'b} \a^-(\lambda ) &=& \a_0^-(\lambda
)\hat{u}_{22,j}^{-}, \qquad
\b^+(\lambda ) = \b_0^+(\lambda ), \\
\label{eq:16.2'c} \c^+(\lambda ) &=& u_{22,j}^{-} \c_0^+(\lambda
), \qquad
\d^+(\lambda ) = u_{22,j}^{-} \d_0^+(\lambda )\hat{u}_{11,j}^{+}, \\
\label{eq:16.2'd} \c^-(\lambda ) &=& \c_0^-(\lambda
)\hat{u}_{11,j}^{+}, \qquad \d^-(\lambda ) = \d_0^-(\lambda ),
\end{eqnarray}\end{subequations}

The dressed solutions of the RHP are given by:
\begin{eqnarray}\label{eq:19.1}
\chi ^+(x,\lambda ) &=& u(x,\lambda )\chi_0 ^+(x,\lambda ) ,\qquad
\chi ^-(x,\lambda ) = u(x,\lambda )\chi_0 ^-(x,\lambda
)\hat{W}(\lambda ) ,
\\
D^+(\lambda ) &=& W(\lambda )D_0^+(\lambda ), \qquad
D^-(\lambda ) = D_0^-(\lambda ) \hat{W}(\lambda ), \\
\label{eq:19.2} W(\lambda ) &=& \openone + (c_ j(\lambda )-1) W_j,
\qquad W_j = \left( \begin{array}{cc} P_{11,j}^{+} & 0 \\ 0 &
P_{22,j}^{-}\end{array} \right).
\end{eqnarray}

The explicit construction of the dressed FAS allow us to reveal
the structure of the FAS at $\lambda \simeq \lambda _j^\pm $. Here
we first formulate the expansions of $\a^\pm(\lambda ) $,
$\c^\pm(\lambda ) $ and their inverse in the vicinities of
$\lambda _j^\pm $
\begin{subequations}\label{eq:29d.1}
\begin{eqnarray}\label{eq:29d.1a}
\a^\pm(\lambda ) &=& \a_{j}^{\pm} +(\lambda -\lambda _j^\pm)
\dot{\a}^\pm(\lambda ) + \mathcal{O}((\lambda -\lambda _j^\pm)^2), \\
\label{eq:29d.1b} \c^\pm(\lambda ) &=& \c_{j}^{\pm} +(\lambda
-\lambda _j^\pm)
\dot{\c}^\pm(\lambda ) + \mathcal{O}((\lambda -\lambda _j^\pm)^2), \\
\label{eq:29d.1c} \hat{\a}^\pm(\lambda ) &=&
{\hat{\a}_{j}^{\pm}\over (\lambda -\lambda _j^\pm)} +
\hat{\dot{\a}}^\pm(\lambda ) + \mathcal{O}((\lambda -\lambda
_j^\pm)), \\
\label{eq:29d.1d} \hat{\c}^\pm(\lambda ) &=& {\hat{\c}_{j}^{\pm}
\over (\lambda-\lambda_j ^\pm)}+ \hat{\dot{\c}}^\pm(\lambda ) +
\mathcal{O}((\lambda -\lambda_j^\pm)),
\end{eqnarray}
\end{subequations}
where
\begin{subequations}\label{eq:29d.2}
\begin{eqnarray}\label{eq:29d.2a}
\a_j^\pm = (\openone -P_{11,j}^{+}) \a_{0,j}^{+}, \qquad
\hat{\a}_j^\pm = (\lambda _j^+-\lambda _j^-)
\hat{\a}_{0,j}^{+}P_{11,j}^{+} , \\
\label{eq:29d.2b} \c_j^\pm = \c_{0,j}^{+} (\openone -P_{11,j}^{+})
, \qquad \hat{\c}_j^\pm = (\lambda _j^--\lambda _j^+) P_{11,j}^{+}
\hat{\c}_{0,j}^{+},
\end{eqnarray}
\end{subequations}

Let us outline the structure of the eigenspaces corresponding to
the discrete eigenvalues $\lambda _j^\pm $. To avoid
technicalities in doing this we will assume that: i)~$L $ has no
other discrete eigenvalues and ii)~that the regular potential
$Q(x) $ is on finite support.  Due to ii) one can prove that all
Jost solutions and their inverse, as well as $T(\lambda ) $ and
$\hat{T}(\lambda ) $ are meromorphic functions of $\lambda $ and can be
extended to the whole complex $\lambda  $-plane. Considering
equations (\ref{eq:6.1}), (\ref{eq:6.2i}) at $\lambda
=\lambda_j^\pm $ we derive the relations:
\begin{subequations}\label{eq:29d.3}
\begin{eqnarray}\label{eq:29d.3a}
|\phi ^\pm_j (x)\rangle  \hat{\a}_j^\pm &=& \pm |\psi ^\pm_j
(x)\rangle \rho _j^\pm , \qquad \rho _j^\pm = \b_j^\pm
\hat{\a}^\pm_j, \\
\label{eq:29d.3b} |\psi ^\pm_j (x)\rangle  \hat{\c}_j^\pm &=& \pm
|\phi ^\pm_j (x)\rangle \tau _j^\pm , \qquad \tau _j^\pm =
\d_j^\mp \hat{\c}^\pm_j,
\end{eqnarray}\end{subequations}
where the index $j $ means that we are taking the value of the
corresponding function for $\lambda =\lambda _j^+ $.

But the eigenfunctions corresponding to the discrete eigenvalues
must be square integrable. A necessary condition for this is the
requirement that these eigenfunctions have no exponentially
growing terms for both $x\to\infty  $ and $x\to-\infty  $. These
limits for $\chi_j ^+(x) \equiv \chi ^+(x,\lambda _j^+) = ( |\phi
_j^+(x)\rangle ,|\psi _j^+(x)\rangle) $ are equal to:
\begin{equation}\label{eq:chi_lim}
\lim_{x\to\infty } \chi _j^+(x) = \re^{-\ri\lambda _j^+Jx} \left(
\begin{array}{cc} \a^+_j & 0 \\ \b^+_j & \openone \end{array}  \right),
\qquad \lim_{x\to -\infty } \chi _j^+(x) = \re^{-\ri\lambda
_j^+Jx} \left(
\begin{array}{cc}\openone   & \d^-_j \\ 0 & \c^+_j\end{array} \right),
\end{equation}
Since generically $\a_j^+\neq 0 $ and $\c_j^+\neq 0 $ some of the
columns of $\chi _j^+(x) $ will be exponentially growing and can
not be interpreted as discrete eigenfunctions. However from eqs.
(\ref{eq:16.2'}) it is clear that both $\a_j^+ $ and $\c_j^+ $ are
degenerate matrices of rank $n-r_j $ and $m-r_j $ respectively.
Thus we find that $r_j $ linear combinations of columns of $|\phi
_j^+(x)\rangle  $ and $|\psi _j^+(x)\rangle  $ due to the
relations
\begin{equation}\label{eq:a_n-0}
\a_j^+ |n_{0,j}^{1}\rangle = 0, \qquad \c_j^+ |n_{0,j}^{2}\rangle
= 0,
\end{equation}
decrease exponentially for both $x\to\infty  $ and $x\to -\infty
$. The eigenspace of $L $ related to $\lambda _j^+ $ is spanned by
$r_j $ linearly independent discrete eigenfunctions which can be
chosen among $|\phi _j^+(x) n_{0,j}^{1}\rangle  $ or $|\psi
_j^+(x) n_{0,j}^{2}\rangle $. These two sets of eigenfunctions
satisfy  linear relations which can be written compactly as:
\begin{equation}\label{eq:16.6'a}
\chi ^+(x,\lambda _j^+)|n_{0,j}\rangle
\mathop{=}\limits^{(\ref{eq:26.1})} (\openone - P_j(x))
|n_{j}(x)\rangle =0.
\end{equation}

The $r_j $ constant vectors $|n_{0,j}\rangle $ which determine the
discrete eigenspace can be viewed as `polarization' vectors of the
corresponding soliton solution and parametrize its internal
degrees of freedom.

The eigenspace corresponding to $\lambda _j^- $ can be analyzed
either independently along the same lines or by using the
reduction properties of $L $, see subsection \ref{ssec:2.2}.
Indeed, from (\ref{eq:16.2'b}) and (\ref{eq:16.2'd}) we get that
only $r_j $ linear combinations of the columns of $|\phi
_j^-(x)\rangle $ (resp. $|\psi _j^-(x)\rangle $) do not have
exponential growth for $x\to\infty  $ (resp. $x\to -\infty $). The
corresponding `polarization' vectors $|n_{0,j}^{+}\rangle  $  are
related to $|n_{0,j}\rangle  $  by:
\begin{equation}\label{eq:n0j-+}
|n_{0,j}^{+}\rangle = D_{0,j}^{+} |n_{0,j}\rangle, \qquad
D_{0,j}^{+}= D_{0}^{+} (\lambda _j^+).
\end{equation}
because
\begin{equation}\label{eq:n0j--}
\a_j^- \c_{0,j}^{+} | n_{0,j}^{2}\rangle =0, \qquad \c_j^-
\a_{0,j}^{+} | n_{0,j}^{1}\rangle =0,
\end{equation}

The analog of eq. (\ref{eq:16.6'a}) for $\chi^-_j(x)$ is the
following relation:
\begin{eqnarray}\label{eq:17.7'}
\chi ^-_j(x)|n_{0,j}^+\rangle &=& 0, \qquad |n_{0,j}^+\rangle =
D_0^+(\lambda _j^+) |n_{0,j}\rangle.
\end{eqnarray}

\begin{remark}\label{rem:3a}
Each of the eigenvalues $\lambda _j^\pm $ corresponds to $r_j
$-dimensional eigensubspace with $1\leq r_j\leq \min (n,m) $. In
particular for the vector NLS (the Manakov model) we may have only
$r_j=1 $. The reduction (\ref{eq:25.1}) relates the sets of
polarization vectors by:
\begin{equation}\label{eq:red-pol}
B^{-1} |m_{0,j}^\dag\rangle =|n_{0,j}\rangle .
\end{equation}

\end{remark}

\subsection{Reflectionless potentials and soliton solutions}
\label{ssec:2.3}

The simplest situation in which the dressing method can be applied
is the one corresponding to vanishing potential and plane wave
solution of (\ref{eq:4.1}):
\begin{equation}\label{eq:Q0}
Q_0(x) = 0, \qquad \chi ^\pm(x,\lambda ) = \re^{-\ri\lambda Jx
-2\ri\lambda ^2Jt}.
\end{equation}
The corresponding scattering matrix of course is $T_0(\lambda
)=\openone $; therefore the corresponding reflection coefficients
(\ref{eq:rho-tau}) are vanishing $\rho ^\pm_0(\lambda )=0 $,
$\tau_0^\pm(\lambda )=0 $ on the whole real $\lambda  $ axis.

Applying the dressing method to (\ref{eq:Q0}) we obtain for the
simplest reflectionless potential:
\begin{eqnarray}\label{eq:Q1}
Q_1(x)= -(\lambda _j^+ - \lambda _j^-) [J,P_j(x,t)].
\end{eqnarray}

Obviously, due to (\ref{eq:16.2'}) the corresponding reflection
coefficients $\rho ^\pm_1(\lambda )=0 $, $\tau_1^\pm(\lambda )=0 $
are also vanishing  on the whole real $\lambda $ axis. Therefore
$Q_1(x) $ is the simplest non-trivial reflectionless potential of
$L $. The scattering matrix for this potential is block-diagonal,
but is not trivial:
\begin{equation}\label{eq:T-rfl}
T_{\rm 1s}(\lambda ) = \left(\begin{array}{cc} u_{11,j}^+(\lambda)
& 0
\\
0 &  \hat{u}_{22,j}^-(\lambda) \\ \end{array} \right)
\end{equation}

The reflectionless potentials are closely related to the soliton
solutions of the corresponding equation. Indeed, the derivation of
the reflectionless potentials we considered the time $t $ as an
auxiliary parameter. In order to go from the reflectionless
potential to the solution of the corresponding NLEE all we need to
do is to impose the correct $t $-dependence on the `polarization
vectors' $|n_{0,j}\rangle $ and $\langle m_{0,j}| $. This can be
determined from the dispersion law of the NLEE as follows:
\begin{equation}\label{eq:nm_t}
|n_{0,j}\rangle \to \exp(-2\ri f(\lambda _j^+)t) |n_{0,j}\rangle,
\qquad \langle m_{0,j}| \to \langle m_{0,j}|\exp(2\ri f(\lambda
_j^-)t).
\end{equation}

Applying this procedure to $Q_1 $ we get the one-soliton solution
of the generic NLEE with dispersion law $f(\lambda ) $ in the
form:
\begin{eqnarray}\label{eq:Q_1s}
Q_{\rm 1s}(x,t)&=& 2(\lambda_j^+ - \lambda_j^-) \\ &\times&\left(
\begin{array}{cc} 0 & -|n_j^1(x,t)\rangle R_j^{-1}(x,t) \langle
m_j^2(x,t)| \\ |n_j^2(x,t)\rangle R_j^{-1}(x,t) \langle
m_j^1(x,t)| & 0
\\
\end{array} \right),\nonumber \\
|n_j^1(x,t)\rangle &=& \re^{-\ri(\lambda_j^+ x + 2 f_{0,j}^+t)}
|n_{0,j}^1\rangle,\qquad |n_j^2(x,t)\rangle = \re^{\ri
(\lambda_j^+ x +
2 f_{0,j}^+t)} |n_{0,j}^2\rangle,\nonumber \\
R_j(x,t)&=& \langle m_j^1(x,t)|n_j^1(x,t)\rangle + \langle
m_j^2(x,t)|n_j^2(x,t)\rangle, \nonumber \\
\langle m_j^1(x,t)| &=& \langle m_{j,0}^1|\re^{\ri (\lambda_j^- x
+ 2 f_{0,j}^-)t)}, \qquad \langle m_j^2(x,t)| = \langle
m_{j,0}^2|\re^{-\ri(\lambda_j^- x + 2 f_{0,j}^-)t)},\nonumber
\end{eqnarray}
where $f_{0,j}^{\pm} =f_0(\lambda _j^\pm) $ and $f(\lambda
)=f_0(\lambda )J $ is the dispersion law  of the NLEE. In order to
obtain the one-soliton solution for the MNLS we have to replace
$f_0(\lambda )J $ by $f_{0,\rm NLS}=-2\lambda ^2$. If we put
$r_j=1$ and take into account the reduction (\ref{eq:25.1}) then
$R_j(x,t)$ can be written down as
\[ R_j(x,t) = 2R_{0,j} \cosh( 2\nu_j x + 2\tilde{\mu}_jt
+\xi_{0,j}),  \] where
\[\nu_j = \ri(\lambda_j^- - \lambda_j^+)/2, \qquad
\widetilde{\mu}_j = \ri(f(\lambda_j^-) - f(\lambda_j^+)),
\]
\[ R_{0,j} = \sqrt{\langle m_{j,0}^1|n_{0,j}^1\rangle
\langle m_{j,0}^2|n_{0,j}^2\rangle}, \qquad \xi_{0,j} = {1\over 2}
\ln {\langle m_{j,0}^1|n_{0,j}^1\rangle \over \langle
m_{j,0}^2|n_{0,j}^2\rangle},\] thus reproducing the well know
result for the one-soliton solution of the vector NLS and the
rank-1 solutions of the MNLS \cite{Ma1,APT}.

We can apply the dressing procedure again, starting with the
potential $Q_{\rm 1s}(x,t) $ and the corresponding FAS $\chi _{\rm
1s}^\pm(x,\lambda ) $. Doing this we have to use a dressing factor
$u_{\rm 2s}(x,\lambda ) $ like in (\ref{eq:26.2}) but with new
locations $\lambda _k^\pm $ for the eigenvalues and a new choice
for the projector $P_k(x) $ which may have different rank $r_k $
and sets of vectors $|n_{k}(x)\rangle $, $\langle m_{k}(x)| $.
Choosing $\lambda _k^\pm \neq \lambda _j^\pm $ the FAS $\chi _{\rm
1s}^\pm(x,\lambda ) $ will be regular at the points $\lambda
=\lambda _k^\pm $ which is all that is required for the procedure
to be valid.

Repeating the procedure several times we get more and more
complicated potentials which are still reflectionless.

Another way to derive these potentials and the related $N
$-soliton solutions of the NLEE consists in using equations
(\ref{eq:X^+}), (\ref{eq:X^-}) with $K^+(x,\lambda )=K^-(x,\lambda
)=0 $. To explain this better following \cite{Ma1} we need to
evaluate the residues of $X^\pm(x,\lambda )\hat{D}^\pm(\lambda ) $
at  $\lambda =\lambda _j^\pm $. Using eq.  (\ref{eq:29d.3}) we
obtain:
\begin{eqnarray}\label{eq:29d.5}
\Res_{\lambda = \lambda _j^+} X^+(x,\lambda)\hat{D}^+(\lambda )
&=& X_j^+(x)\mathcal{K}_j^+(x), \\
\label{eq:29d.5m} \Res_{ \lambda =\lambda _j^-}
X^-(x,\lambda)\hat{D}^-(\lambda ) &=& -X_j^-(x)
\mathcal{K}_j^-(x),
\end{eqnarray}
where $X_j^\pm(x) =X^\pm(x,\lambda _j^\pm) $ and
\begin{eqnarray}\label{eq:29d.K}
\mathcal{K}_j^+(x) = \left( \begin{array}{cc} 0 & \rho_j^+ (x)  \\
\tau_j^+ (x)  & 0 \end{array}\right), &\qquad & \mathcal{K}_j^-(x)
= \left( \begin{array}{cc} 0 & \tau _j^-(x) \\ \rho_j^-(x) & 0
\end{array}\right), \\
\rho _j^\pm (x) = \rho _j^\pm \re^{\pm 2\ri\lambda _j^\pm x}, &
\qquad & \tau _j^\pm (x) = \tau _j^\pm \re^{\mp 2\ri\lambda _j^\pm
x}. \nonumber
\end{eqnarray}
Thus the equations (\ref{eq:X^+}), (\ref{eq:X^-}) take the form:
\begin{eqnarray}\label{eq:*-2}
X^+(x,\lambda ) &=&  \openone  - \sum_{j=1}^{N} {1 \over \lambda
-\lambda
_j^-} X_j^-(x) \mathcal{K}_j^-(x), \\
\label{eq:*-3} X^-(x,\lambda ) &=&  \openone  + \sum_{j=1}^{N} {1
\over \lambda -\lambda _j^+} X_j^+(x) \mathcal{K}_j^+(x).
\end{eqnarray}
Note that it is legitimate to put $\lambda =\lambda _k^+ $ in eq.
(\ref{eq:*-2}) and $\lambda =\lambda _k^- $ in eq. (\ref{eq:*-3})
thus getting an algebraic system of equations for $X_j^\pm(x) $.
After solving for $X_j^\pm(x) $ we can insert them into eqs.
(\ref{eq:X^+}), (\ref{eq:X^-}) thus recovering the FAS
$X^\pm(x,\lambda ) $ for all $\lambda \in \bbbc_\pm $. The
corresponding reflectionless potential is given by (see eq.
(\ref{eq:DR2.5})):
\begin{equation}\label{eq:*-4}
Q^{(N)}(x) = - \sum_{j=1}^{N} [J, X_j^-(x)\mathcal{K}_j^-(x)] =
\sum_{j=1}^{N} [J, X_j^+(x)\mathcal{K}_j^+(x)] .
\end{equation}

As we shall see in the next subsection the generalized
Zakharov-Shabat system (\ref{eq:4.1}) with potential $Q^{(N)}(x) $
has $2N $ discrete eigenvalues located at $\lambda _j^\pm $ with
$r_j $-dimensional eigenspaces. The corresponding $N $-soliton
solution $Q_{\rm Ns}(x,t) $ is obtained from (\ref{eq:*-4}) by
introducing the relevant $t $-dependence into the polarizations
vectors $|n_{0,j}\rangle  $ and $\langle m_{0,j}| $ or
equivalently, in $\mathcal{K}_j^\pm(x,t) $ as follows:
\begin{equation}\label{eq:*-5}
\rho_j^\pm (x,t) = \rho _j^\pm(x) \re^{\pm 2\ri f_{0,j}^\pm t},
\qquad \tau_j^\pm (x,t) = \tau_j^\pm(x) \re^{\mp 2\ri f_{0,j}^\pm
t},
\end{equation}
where $f_{0,j}^{\pm} = f_0(\lambda _j^\pm) $ and $f(\lambda
)=f_0(\lambda )J $ is the dispersion law of the  NLEE.

\subsection{Spectral properties of $L $}\label{ssec:2.2sp}

Using the FAS one can construct the resolvent of (\ref{eq:4.1}):
\begin{eqnarray}\label{eq:7.2}
R^\pm(x,y,\lambda ) &=& -\ri \chi ^\pm(x,\lambda ) \Theta ^\pm
(x-y)
\hat{\chi }^\pm(y,\lambda ), \\
\Theta^\pm (z)&=& \diag(\mp \theta(\mp z)\openone  , \pm
\theta(\pm z) \openone ), \nonumber
\end{eqnarray}
where $\theta (z) $ is the standard step-function.

Let us consider $R^\pm(x,y,\lambda ) $ as the kernel of an
integral operator acting on vector-valued functions of $x $ as
follows:
\begin{equation}\label{eq:res}
(\R_\lambda  f)(x) = \int_{-\infty }^{\infty } \rd y
R^\pm(x,y,\lambda ) f(y), \qquad \mbox{for $\lambda \in \bbbc_\pm
$}.
\end{equation}

\begin{theorem}\label{th:R2}
Let $Q(x) $ satisfy conditions (C.1) and (C.2) and let $\lambda
_j^\pm $ be the zeroes of $\det\a^\pm(\lambda ) $. Then
\begin{enumerate}

\item $R^\pm (x,y,\lambda ) $ is an analytic function of $\lambda
$ for $\lambda \in \bbbc_\pm $ having pole singularities at
$\lambda _j^\pm \in \bbbc_\pm $;

\item $R^\pm (x,y,\lambda ) $ is a kernel of a bounded integral
operator for $\im \lambda \neq 0 $;

\item $R^\pm (x,y,\lambda ) $ is uniformly bounded function for
$\lambda \in\bbbr $ and provides a kernel of an unbounded integral
operator;

\item $R^\pm (x,y,\lambda ) $ satisfy the equation:
\begin{equation}\label{eq:R3.1}
L(\lambda ) R^\pm (x,y,\lambda )=\openone \delta (x-y).
\end{equation}
\end{enumerate}
\end{theorem}

\begin{proof}[Idea of the proof] {}

\begin{enumerate}

\item is obvious from the fact that $\chi ^\pm(x,\lambda ) $ are
the FAS of $L(\lambda ) $. From the definition (\ref{eq:6.3})
there follows that $\det \chi ^\pm(x,\lambda )=\det \a^\pm(\lambda
) $, i.e. $\hat{\chi }^\pm(y,\lambda ) $ and consequently,
$R^\pm(x,y,\lambda ) $ will develop pole singularities for all
$\lambda _j^\pm $ for which $\det\a^\pm(\lambda )=0$.

\item Assume that $\im \lambda >0 $ and consider the asymptotic
behavior of $R^+ (x,y,\lambda ) $ for $x,y\to\infty  $. From
equations (\ref{eq:6.3}), (\ref{eq:6.4}) we find that
\begin{eqnarray}\label{eq:R3.2}
R^+ (x,y,\lambda ) &=& \sum_{p=1}^{n} X^+(x,\lambda )
\re^{-\ri\lambda J(x-y)} \Theta^+(x-y) \hat{X}^+(y,\lambda ) \\
\nonumber
\end{eqnarray}

Due to the fact that $\chi ^+(x,\lambda ) $ has block-triangular
asymptotics for $x\to\infty  $ and $\lambda \in\bbbc_+ $ and for
the correct choice of $\Theta^+(x-y) $ (\ref{eq:7.2}) we check
that the right hand side of (\ref{eq:R3.2}) falls off
exponentially for $x\to\infty  $ and arbitrary choice of $y $. All
other possibilities are treated analogously.

\item For $\lambda \in\bbbr$ the arguments of item 2) can not be
applied because the exponentials in the right hand side of
(\ref{eq:R3.2}) $\im \lambda =0  $ only oscillate. Thus we
conclude that $R^\pm(x,y,\lambda ) $ for $\lambda \in\bbbr $ is
only a bounded function and thus the corresponding operator
$R(\lambda ) $ is an unbounded integral operator.

\item The proof of eq. (\ref{eq:R3.1}) follows from the fact that
$L(\lambda )\chi ^+(x,\lambda )=0 $ and
\begin{equation}\label{eq:R4.1}
{\rd\Theta^\pm (x-y)  \over \rd x } = \openone \delta (x-y).
\end{equation}
\end{enumerate}
The theorem is proved.
\end{proof}

From theorem \ref{th:R2}, item 3) there follows that the
continuous spectrum of $ L $  fills up the whole real $\lambda
$-axis with multiplicity $n+m $. By definition the operator $L $
may also have discrete eigenvalues at the points at which
$R^\pm(x,y,\lambda ) $ have pole singularities. From item 1) it
follows that these are precisely the points $\lambda _j^\pm $.

Let us now analyze the structure of these singularities and
evaluate the corresponding residues. To this end we insert eqs.
(\ref{eq:19.1}), (\ref{eq:19.2}) into (\ref{eq:7.2}). The result
is that $R^\pm(x,y,\lambda )$ have poles of first order in the
neighborhood  of $\lambda_j^\pm$ with residues :
\begin{equation}\label{eq:28.2}
\Res_{\lambda =\lambda _j^+} R^+(x,y,\lambda ) = -\ri (\lambda
_j^+-\lambda _j^-) (\openone -P_j(x))\chi _{0,j}^+(x) \Theta
^+(x-y) \hat{\chi }_{0,j}^+(y) P_j(y).
\end{equation}
\begin{equation}\label{eq:28.4}
\Res_{\lambda =\lambda _j^-} R^-(x,y,\lambda ) = \ri (\lambda
_j^+-\lambda _j^-) P_j(x)\chi _{0,j}^-(x) \Theta ^-(x-y) \hat{\chi
}_{0,j}^-(y)(\openone - P_j(y)).
\end{equation}
Formally in the right hand side of (\ref{eq:28.2}) there enter the
discontinuous functions $\Theta ^\pm(x-y) $. However, due to the
special structure of the projectors $ P_j(x)$ (see eqs.
(\ref{eq:26.3}) and (\ref{eq:29d.3a}), (\ref{eq:29d.3b}) we obtain
the following expressions for the residues of $R^\pm(x,y,\lambda
)$:
\begin{equation}\label{eq:28.6}
\Res_{\lambda =\lambda _j^\pm} R^\pm(x,y,\lambda ) = \pm \ri |\psi
_j^\pm (x)\rangle \rho _j^\pm \langle \psi _j^\pm (y)|.
\end{equation}
where $\rho _j^\pm $ are defined in eq. (\ref{eq:29d.3}) and there
are no discontinuities present.

Now we can derive the completeness relation for the eigenfunctions
of the Lax operator (\ref{eq:4.1}) by applying the contour
integration method (see e.g. \cite{GeKu,AKNS}) to the integral:
\begin{equation}\label{eq:3.38}
\mathcal{J}(x,y)={1\over 2\pi \ri}\oint_{\gamma _+} \rd\lambda
R^{+}(x,y,\lambda)- {1\over 2\pi \ri}\oint_{\gamma _-} \rd\lambda
R^{-}(x,y,\lambda),
\end{equation}
where the contours $\gamma _\pm$ are shown on the Figure
\ref{fig:1}. Skipping the details we get:
\begin{equation}\label{eq:3.44}
\begin{split}
&\delta(x-y) J\\ &= {1\over 2\pi}\int_{-\infty}^\infty
\rd\lambda\left\{ |\phi^+(x,\lambda)\rangle \hat{\a}^+(\lambda )
\langle \hat{\psi}^{+}(y,\lambda)| - |\phi^{-}(x,\lambda)\rangle
\hat{\a}^-(\lambda )\langle \hat{\psi}^{-}(y,\lambda)|\right\} \\
&-\ri \sum_{j=1}^N \left( |\psi _j^+ (x)\rangle \rho _j^+ \langle
\psi _j^+ (y)| - |\psi _j^- (x)\rangle \rho _j^- \langle \psi _j^-
(y)|\right).
\end{split}
\end{equation}

The completeness relation (\ref{eq:3.44}) is a natural
generalization of the  one in  \cite{GeKu} for the $sl(2)$ case.
An important difference here is that now we have matrix-valued
spectral functions $\a^\pm(\lambda)$ whose zeroes determine the
location of the discrete eigenvalues.

\begin{remark}\label{rem:4}
When both $n>1 $ and $m>1 $ there are two possible definitions of
simple eigenvalues. One of them used in \cite{APT} is to define
$\lambda _j^\pm$ as simple if $\det \a^\pm(\lambda ) $ have simple
zeroes for $\lambda =\lambda _j^\pm $.  The other possible
definition is:  the eigenvalues $\lambda _j^\pm $ are simple if
the resolvent of $L $ has simple poles at $\lambda _j^\pm $. The
singular solutions of the RHP (\ref{eq:26.2}) correspond to simple
poles of the resolvent (\ref{eq:7.2}) although in the neighborhood
of $\lambda \simeq \lambda _j^\pm $ we have that $\det \a^\pm(\lambda
)$ behaves like $ (\lambda -\lambda _j^\pm)^{r_j} $.

\end{remark}

\section{The Generalized Fourier Transforms for Nonregular $J$
}\label{ch:GFT}

The main result in this section consists in the fact that the
analysis of \cite{G} can be applied also to nonregular choices of
$J$. In addition we briefly outline also how to take into account
the presence of discrete spectrum of $L$. Skipping the details of
the proof we formulate the completeness relation for the `squared'
solutions of $L$.

\subsection{The Wronskian relations}\label{sec:l5-1}

The analysis of the mapping $\mathcal{ F} \colon \mathcal{ M} \to
\mathcal{ T} $ between the class of allowed potentials $\mathcal{
M} $ and the scattering data of $L $ starts with the Wronskian
relations, see \cite{CaDe,CaDeB} for $sl(2) $-case and
\cite{TMF98} for the block-matrix case (\ref{eq:4.1}).  As we
shall see, they would allow us to

\begin{enumerate}

\item  To formulate the idea that the ISM is a GFT;

\item To determine explicitly the proper generalizations of the
usual exponents;

\item To introduce the skew--scalar product on $\mathcal{ M} $
which provides it with a symplectic structure.

\end{enumerate}

All these ideas  will be worked out for the system (\ref{eq:4.1}).
With (\ref{eq:4.1}) one can associate the systems:
\begin{eqnarray}\label{eqB:I.3}
&& \ri {\rd\hat{\psi} \over \rd x }- \hat{\psi} (x,t,\lambda
)U(x,t,\lambda ) =0, \qquad U(x,\lambda )=Q(x)-\lambda J, \\
\label{eqB:I.4} && \ri {\rd \delta \psi \over \rd x }  + \delta
U(x,t,\lambda ) \psi
(x,t,\lambda ) + U(x,t,\lambda ) \delta \psi (x,t,\lambda ) =0\\
\label{eqB:I.5} && \ri {\rd\dot{\psi} \over \rd x }  - \lambda J
\psi (x,t,\lambda ) + U(x,t,\lambda ) \dot{\psi} (x,t,\lambda ) =0
\end{eqnarray}
where $\delta \psi  $ corresponds to a given variation $\delta
Q(x,t) $ of the potential, while by dot we denote the derivative
with respect to the spectral parameter.

We start with the identity:
\begin{eqnarray}\label{eqB:wr.1}
\left. \left( \hat{\chi }J \chi (x,\lambda ) - J \right)
\right|_{x=-\infty }^{\infty } &=& - \ri\int_{-\infty }^{\infty }
\rd x\, \hat{\chi }[Q(x), J]\chi (x,\lambda ),
\end{eqnarray}
where $\chi (x,\lambda ) $ can be any fundamental solution of $L
$. For convenience we choose them to be the FAS introduced above.

The l.h.side of (\ref{eqB:wr.1}) can be calculated explicitly by
using the asymptotics of $\chi ^\pm(x,\lambda ) $ for $x\to \pm
\infty  $, (\ref{eq:6.3}), (\ref{eq:6.4}).  It would be expressed
by the matrix elements of the scattering matrix $T(\lambda ) $,
i.e., by the scattering data of $L $ as follows:
\begin{subequations}\label{eq:21.2}
\begin{eqnarray}\label{eqB:wr.2}
\left. \left( \hat{\chi }^+J \chi^+ (x,\lambda ) - J \right)
\right|_{x=-\infty }^{\infty } &=& -2 \left( \begin{array}{cc}
0 & \d^-(\lambda )\\ \b^+(\lambda ) & 0 \end{array} \right)\\
\label{eqB:wr.3} \left. \left( \hat{\chi }^-J \chi^- (x,\lambda )
- J \right) \right|_{x=-\infty }^{\infty } &=& -2 \left(
\begin{array}{cc} 0 & \b^-(\lambda )\\ \d^+(\lambda ) & 0
\end{array} \right)
\end{eqnarray}
\end{subequations}

We will show that these Wronskian relations allow us to express
the elements of each of the sets $\mathcal{T}_i $, $i=1,2 $ in eq.
(\ref{eq:T_12}) as integrals from the potential $Q(x) $ multiplied
by some bilinear combination of eigenfunctions of $L $ called
`squared solutions'.  Let us now analyze the equations obtained
from (\ref{eqB:wr.1}) after multiplying by the matrix $E_{ab} $,
$(E_{ab})_{cd}= \delta _{ac}\delta _{bd} $ and taking the trace.
Such operation will produce the $a,b $-matrix element of eq.
(\ref{eqB:wr.1}). In the right hand side of this equation we can
use the invariance properties of the trace and rewrite it in the
form:
\begin{eqnarray}\label{eqB:wr.1ab}
\tr \left(\left. \left( \hat{\chi }J \chi (x,\lambda ) - J
\right)E_{ab}\right)\right|_{x=-\infty }^{\infty }&=& -
\ri\int_{-\infty }^{\infty } \rd x\, \tr \left( [Q(x), J]
\e_{ab}(x,\lambda )\right),
\end{eqnarray}
where
\begin{equation}\label{eq:e-ab}
e_{ab}(x,\lambda )=\chi E_{ab}\hat{\chi } (x,\lambda )  ,\qquad
\e_{ab}(x,\lambda )=P_{0J}(\chi E_{ab}\hat{\chi } (x,\lambda ))  ,
\end{equation}
are the natural generalization of the `squared solutions'
introduced first for the $sl(2) $-case \cite{KN79,DJK1}. By
$P_{0J}$ we have denoted the projector $P_{0J}=\ad_J^{-1}\ad_J$ on
the block-off-diagonal part of the corresponding matrix-valued
function.

The `squared solutions' obviously satisfy the equation:
\begin{equation}\label{eq:e-ab1}
\ri {\rd e_{ab}\over \rd x } +[Q(x)-\lambda J,e_{ab}(x,\lambda
)]=0.
\end{equation}
The indices $a,b $ in (\ref{eq:e-ab})  are taking values in
appropriate ranges; for convenience below $i,h,k $ (resp. by
$l,r,s $)  will be taking values in the ranges:
\begin{equation}\label{eq:range}
1 \leq i,h,k \leq n, \qquad n+1 \leq l,r,s \leq n+m.
\end{equation}
and by $i<r$ we will mean $1\leq i\leq n$ and $n+1\leq r\leq n+m$.
We also introduce:
\begin{eqnarray}\label{eq:40-41}
\begin{split}
\Psi _{ab}^{\pm} (x,\lambda ) &=& |\psi ^\pm \rangle \epsilon
_{ab} \langle \hat{\psi }^\pm| ,\qquad \Phi _{ab}^{\pm} (x,\lambda
) = |\phi ^\pm \rangle \epsilon _{ab} \langle \hat{\phi }^\pm| ,
\\
\Theta _{ab}^{\pm} (x,\lambda ) &=& |\phi ^\pm \rangle \epsilon
_{ab} \langle \hat{\psi }^\pm|, \qquad \Xi _{ab}^{\pm} (x,\lambda
) = |\psi ^\pm \rangle \epsilon _{ab} \langle \hat{\phi}^\pm| ,
\end{split}
\end{eqnarray}
where by $\epsilon _{ab} $ we mean the relevant non-vanishing
blocks of the matrices $E_{ab} $; e.g.:
\begin{equation}\label{eq:E-ab}
\begin{split}
E_{ih} = \left( \begin{array}{cc} \epsilon _{ih} & 0 \\ 0 & 0
\end{array}
\right), \qquad E_{rs} = \left( \begin{array}{cc} 0 & 0 \\ 0 &
\epsilon _{rs}\end{array}
\right), \\
E_{ir} = \left( \begin{array}{cc} 0 & \epsilon _{ir} \\ 0 & 0
\end{array}
\right), \qquad E_{ri} = \left( \begin{array}{cc} 0 & 0 \\
\epsilon _{ri} &0
\end{array}
\right),
\end{split}
\end{equation}

Using these definitions and eqs. (\ref{eq:6.3}), (\ref{eq:6.4}) we
find that the `squared solutions' can be expressed in terms of the
Jost solutions as follows:
\begin{eqnarray}\label{eq:e-Psi}
\begin{split}
e_{li}^+ (x,\lambda )= \hat{a}^+_{ih}(\lambda ) \Psi ^+_{lh}
(x,\lambda ), \qquad e_{il}^+ (x,\lambda )= \hat{c}^+_{lr}(\lambda
) \Phi ^+_{ir}(x,\lambda ),
\\
e_{ih}^+ (x,\lambda )= \hat{a}^+_{hk}(\lambda )\Theta
^+_{ik}(x,\lambda ), \qquad e_{lr}^+ (x,\lambda )=
\hat{c}^+_{rs}(\lambda ) \Xi ^+_{ls}(x,\lambda ).
\end{split} \\
\label{eq:e-Phi}
\begin{split}
e_{il}^- (x,\lambda )= \hat{a}^-_{lr}(\lambda ) \Psi ^-_{ir}
(x,\lambda ), \qquad e_{li}^- (x,\lambda )=
\hat{c}^-_{ih}(\lambda ) \Phi ^-_{lh}(x,\lambda ),
\\
e_{ih}^- (x,\lambda )= \hat{c}^-_{hk}(\lambda )\Theta
^-_{ik}(x,\lambda ), \qquad e_{rs}^- (x,\lambda )=
\hat{a}^+_{sl}(\lambda ) \Xi ^-_{rl}(x,\lambda ).
\end{split}
\end{eqnarray}
Here we assume summation over the repeated indices in the relevant
range (\ref{eq:range}).

We will also need the skew--scalar product:
\begin{equation}\label{skew}
\biglb X, Y \bigrb = {1 \over 2}\int_{-\infty}^\infty \rd x \tr
\left( X(x), [J, Y(x) ] \right),
\end{equation}
which is non-degenerate for $ X(x), Y(x) \in \mathcal{M} $. Then
we can write down the reflection coefficients $\rho _{ab}^{\pm} $
in the form:
\begin{equation}\label{eq:rho-ab}
\rho _{lk}^{+}(\lambda ) = -\ri \biglb Q(y),e_{il}^{+}(y,\lambda
)\bigrb \hat{a}^+_{ik} ,
\end{equation}
and similar expressions for the other reflection coefficients.
Thus we have a formula analogous to the standard Fourier transform
in which $e_{il}^{+}(y,\lambda ) $ can be viewed as the
generalizations of the standard exponentials.

In order to work out the contributions from the discrete spectrum
of $L $ we will need the first two coefficients in the Taylor
expansions of $\Psi ^\pm_{ab}(x,\lambda ) $ and $\Phi
^\pm_{ab}(x,\lambda ) $;
\begin{equation}\label{eq:Psi-exp}
\begin{split}
\Psi ^\pm_{ab}(x,\lambda ) = \Psi ^\pm_{ab;j}(x) + (\lambda
-\lambda _j^\pm ) \dot{\Psi} ^\pm_{ab;j}(x) + \mathcal{O}((\lambda
-\lambda
_j^\pm)^2), \\
\Phi ^\pm_{ab}(x,\lambda ) = \Phi ^\pm_{ab;j}(x) + (\lambda
-\lambda _j^\pm ) \dot{\Phi} ^\pm_{ab;j}(x) + \mathcal{O}((\lambda
-\lambda _j^\pm)^2),
\end{split}
\end{equation}
where:
\begin{equation}\label{eq:not}
\Phi_{ab;j}^{\pm} = \Phi_{ab}^{+}(x,\lambda _j^\pm), \qquad
\dot{\Phi}_{ab;j}^{\pm} =\left. {\rd \Phi_{ab}^{+}(x,\lambda
)\over \rd \lambda } \right|_{\lambda = \lambda _j^\pm}.
\end{equation}

Indeed, from eq. (\ref{eq:40-41}) there follows that these are
regular functions for all $\lambda\in \bbbc_\pm $. However the
`squared solutions' $e_{ab}^{\pm}(x,\lambda ) $ have pole
singularities of first order in the vicinities of $\lambda _j^\pm
$ and therefore we will have
\begin{equation}\label{eq:e-exp}
e^\pm_{ab}(x,\lambda ) \simeq  {e^\pm_{ab;j}(x) \over (\lambda
-\lambda _j^\pm )} + \dot{e} ^\pm_{ab;j}(x) + \mathcal{O}(\lambda
-\lambda_j^\pm),
\end{equation}
where for $a=i $ and $b=l $ we have:
\begin{equation}\label{eq:e-coef}
e^+_{il;j}(x) = \hat{\a}_{ih;j}^+ \Psi _{lh;j}^+(x), \qquad
\dot{e}^+_{il;j}(x) = \hat{\dot{\a}}_{ih;j}^+ \Psi _{lh;j}^+(x) +
\hat{\a}_{ih;j}^+ \dot{\Psi} _{lh;j}^+(x),
\end{equation}
and similar expressions for $e^\pm_{ab;j}(x) $ and
$\dot{e}^\pm_{ab;j}(x) $ with different choices for the indices
$a,b $.

The second type of Wronskian relations, which we will consider
relate the variation of the potential $\delta Q(x) $ to the
corresponding variations of the scattering data. To this purpose
we start with the identity:
\begin{eqnarray}\label{eqB:wr.16}
\left. \hat{\chi }\delta  \chi (x,\lambda ) \right|_{x=-\infty
}^{\infty } &=& -\ri \int_{-\infty }^{\infty } \rd x\, { \rd \over
\rd x} \left( i\hat{\chi }\delta  \chi \right)(x,\lambda ).
\end{eqnarray}
To calculate the integrand in (\ref{eqB:wr.16}) we need to use
the equation (\ref{eqB:I.4}), satisfied by $\delta \chi (x,\lambda
) $. Making use of eqs. (\ref{eq:4.1}) and (\ref{eqB:I.4}) we get:
\begin{eqnarray}\label{eqB:wr.18}
\left. \hat{\chi }\delta  \chi (x,\lambda ) \right|_{x=-\infty
}^{\infty } = \ri\int_{-\infty }^{\infty } \rd x\, \hat{\chi
}\delta Q(x)\chi (x,\lambda ).
\end{eqnarray}

We apply ideas similar to the ones above. Evaluating the l.h.side
of (\ref{eqB:wr.18}) with $\chi (x,\lambda )\equiv \chi^+
(x,\lambda ) $ and $\chi (x,\lambda )\equiv \chi^- (x,\lambda ) $
we find:
\begin{eqnarray}\label{eqB:wr.19}
\left. \hat{\chi }^+\delta  \chi^+(x,\lambda)\right|_{x=-\infty
}^{\infty } = \left( \begin{array}{cc} \hat{\a}^+ \delta
\a^+(\lambda ) & - \delta \tau^+(\lambda ) \c^+(\lambda ) \\
\delta \rho ^+(\lambda )\a^+(\lambda ) & - \hat{\c}^+ \delta
\c^+(\lambda ) \end{array} \right),\\
\label{eqB:wr.20} \left. \hat{\chi }^-\delta
\chi^-(x,\lambda)\right|_{x=-\infty }^{\infty } = \left(
\begin{array}{cc} \hat{\c}^- \delta \c^-(\lambda )   &
 \delta \rho^-(\lambda ) \a^-(\lambda ) \\ -\delta \tau^-(\lambda )
\c^-(\lambda ) & - \hat{\a}^-(\lambda )\delta \a^-(\lambda ) .
\end{array} \right),
\end{eqnarray}

Multiplying by $E_{ab}$ and taking the trace we arrive at:
\begin{equation}\label{eq:drho-ab}
\begin{split}
\delta \rho _{lk}^{+}(\lambda ) &= 2\ri \biglb
e_{il}^{+}(y,\lambda),\ad_J^{-1} \delta Q(y)\bigrb \hat{a}^+_{ik},
\\ \delta \rho _{ik}^{-}(\lambda ) &=- 2\ri \biglb
e_{li}^{+}(y,\lambda), \ad_J^{-1} \delta Q(y)\bigrb \hat{a}^+_{lk}
, \end{split}
\end{equation}

These relations are basic in the analysis of the related NLEE and
their Hamiltonian structures. Below we shall use them assuming
that the variation of $Q(x) $ is due to its time evolution, and
consider variations of the type:
\begin{eqnarray}\label{eqB:wr.23}
\delta  Q(x,t) = Q_t \delta t + \mathcal{ O} ((\delta  t)^2).
\end{eqnarray}
Keeping only the first order terms with respect to $\delta t $ we
find:
\begin{eqnarray}\label{eqB:wr.24}
\begin{split}
{\rd\rho _{lk}^{+}(\lambda )\over \rd t} &= 2\ri \biglb
e_{il}^{+}(y,\lambda), \ad_J^{-1} Q_t(y)\bigrb \hat{a}^+_{ik}, \\
{\rd\rho _{ik}^{-}(\lambda )\over \rd t} &=- 2\ri \biglb
e_{li}^{-}(y,\lambda), Q_t(y)\bigrb \hat{a}^-_{lk} .
\end{split}
\end{eqnarray}

\subsection{Completeness of the `squared solutions' }
Let us introduce the sets of `squared solutions'
\begin{eqnarray}\label{eq:Psi-}
\{\bPsi \} = \{\bPsi \}_{\rm c} \cup \{\bPsi \}_{\rm d}, \qquad
\{\bPhi \} = \{\bPhi \}_{\rm c} \cup \{\bPhi \}_{\rm d},\\
\label{eq:Psi}
\begin{split}
\{\bPsi \}_{\rm c} &\equiv \left\{ \bPsi ^+_{ri}(x,\lambda), \quad
\bPsi ^-_{ir}(x,\lambda), \quad i<r, \quad \lambda \in \bbbr
\right\},
\\
\{\bPsi \}_{\rm d} &\equiv \left\{\bPsi ^+_{ri;j}(x), \quad
\dot{\bPsi }^+_{ri;j}(x),\quad  \bPsi ^-_{ir;j}(x),\quad
\dot{\bPsi }^-_{ir;j}(x)\right\}_{j=1}^{N},
\end{split}\\
\label{eq:Phi}
\begin{split}
\{\bPhi \}_{\rm c} &\equiv \left\{ \bPhi ^+_{ir}(x,\lambda), \quad
\bPhi ^-_{ri}(x,\lambda), \quad i<r, \quad \lambda \in \bbbr
\right\},
\\
\{\bPhi \}_{\rm d} &\equiv \left\{\bPhi ^+_{ir;j}(x), \quad
\dot{\bPhi }^+_{ir;j}(x),\quad  \bPhi ^-_{ri;j}(x),\quad
\dot{\bPhi }^-_{ri;j}(x)\right\}_{j=1}^{N},
\end{split}
\end{eqnarray}
where the subscripts `c' and `d' refer to the continuous and
discrete spectrum of $L $. The `squared solutions' in bold-face
$\bPsi_{ri}^{+} $, \dots are obtained from $\Psi _{ri}^{+} $, \dots by
applying the projector $P_{0J} $, i.e. $\bPsi _{ri}^{+}(x,\lambda ) =
P_{0J}\Psi _{ri}^{+}(x,\lambda ) $, see eq. (\ref{eq:e-ab}).

\begin{theorem}[see \cite{TMF98}]\label{t2.1}
The sets $\{\bPsi \} $  and $\{\bPhi \} $ form complete sets of
functions in $\mathcal{M}_J$. The corresponding completeness
relation has the form:
\begin{eqnarray}\label{eq:5.23}
\begin{split}
\delta(x-y)\Pi_{0J} &= {1\over \pi} \int_{-\infty}^\infty
\rd\lambda
(G^+(x,y,\lambda) - G^-(x,y,\lambda) ) \\
&- 2\ri \sum_{j=1}^{N} (G_j^+(x,y) + G_j^-(x,y) ),
\end{split}
\end{eqnarray}
where
\begin{eqnarray}\label{eq:5.23'}
\Pi_{0J} =\sum_{i<r}( E_{ir}\otimes E_{ri} - E_{ri}\otimes E_{ir}) ,\\
\begin{split}
G^+(x,y,\lambda) &= \sum_{i<r}\e_{ir}^+(x,\lambda)\otimes
\e_{ri}^+(y,\lambda),\\
G^-(x,y,\lambda) &= \sum_{i<r} \e_{ri}^-(x,\lambda)\otimes
\e_{ir}^-(y,\lambda),
\end{split}\\
\begin{split}
G_j^+(x,y) &= \sum_{i<r} (\e_{ir;j}^+(x)\otimes
\dot{\e}_{ri;j}^+(y) + \dot{\e}_{ir;j}^+(x)\otimes \e_{ri;j}^+(y)), \\
G_j^-(x,y) &= \sum_{i<r}
(\dot{\e}_{ri;j}^-(x)\otimes\e_{ir;j}^-(y)
+\e_{ri;j}^-(x)\otimes\dot{\e}_{ir;j}^-(y)),
\end{split}\end{eqnarray}
\end{theorem}

\begin{proof}[Idea of the proof:]
Apply the contour integration method to a conveniently chosen
Green function, see \cite{TMF98}.
\end{proof}

\subsection{Expansions over the ,,squared'' solutions}\label{sec:l5-3}

Using the completeness relations one can expand any generic
element $F(x) $ of the phase space $ \mathcal{ M} $ over each of
the complete sets of `squared solutions'. We remind that $F(x) $
is a generic element of $\mathcal{ M} $ if it is a
block-off-diagonal matrix-valued function, which falls off fast
enough for $|x|\to\infty  $. It can be written down in terms of
its matrix elements $F_\pm(x) $ as:
\begin{eqnarray}\label{eqB:4.4a}
F(x) = \sum_{i<r}( F_{ir}(x) E_{ir} + F_{ri}(x) E_{ri}).
\end{eqnarray}
From (\ref{eq:5.23'}) we get:
\begin{eqnarray}\label{eqB:4.4b}
-{1  \over 2 } \tr_1 \left([J, F(x)] \otimes \openone \right)
\Pi_{0J} = {1  \over 2 } \tr_2 \Pi_{0J} \left(\openone \otimes [J,
F(x)] \right)  =  F(x).
\end{eqnarray}
where $\tr_1 $ (and $\tr_2 $) mean that we are taking the trace of
the elements in the first (or the second) position of the tensor
product. The result is:
\begin{eqnarray}\label{3.5}\begin{split}
F(x) &= {1 \over \pi} \int_{-\infty}^\infty d\lambda \sum_{i<r}
\left(\e^+_{ir}(x,\lambda) \gamma^+_{F;ir}(\lambda) -
\e^-_{ri}(x,\lambda) \gamma^-_{F;ri}(\lambda) \right)\\ &-2\ri
\sum_{j=1 }^N( Z_{F;j}^+(x) +  Z_{F;j}^-(x)),\\
\end{split}\\
\begin{split}
F(x) &= -{1 \over \pi} \int_{-\infty}^\infty d\lambda \sum_{i<r}
\left(\e^+_{ri}(x,\lambda) \gamma^+_{F;ri}(\lambda) -
\e^-_{ir}(x,\lambda) \gamma^-_{F;ir}(\lambda) \right) \\ &-2\ri
\sum_{j=1 }^N( \tilde{Z}_{F;j}^+(x) +  \tilde{Z}_{F;j}^-(x)),
\end{split}
\end{eqnarray}
where
\begin{eqnarray}\label{3.5'}
\gamma^\pm _{F;ab}(\lambda) = \biglb \e^\pm_{ba}(y,\lambda),
F(y)\bigrb, \\ \label{3.5''}
\begin{split}
Z_{F;j}^+(x) = \Res_{\lambda=\lambda_j^+} \sum_{i<r}
\e^+_{ir}(x,\lambda) \gamma^+_{F;ir}(\lambda), \\ Z_{F;j}^-(x) =
\Res_{\lambda=\lambda_j^-} \sum_{i<r} \e^-_{ri}(x,\lambda)
\gamma^-_{F;ri}(\lambda),
\end{split}\\ \label{3.5'''}
\begin{split}
\tilde{Z}_{F;j}^+(x) = \Res_{\lambda=\lambda_j^+} \sum_{i<r}
\e^+_{ri}(x,\lambda) \gamma^+_{F;ri}(\lambda), \\
\tilde{Z}_{F;j}^-(x) = \Res_{\lambda=\lambda_j^-} \sum_{i<r}
\e^-_{ir}(x,\lambda) \gamma^-_{F;ir}(\lambda),
\end{split}
\end{eqnarray}

The completeness relation (\ref{eq:5.23}) is directly related to
the spectral decompositions of $\Lambda_\pm$, see e.g.
\cite{LMP,GeKu2} for $\frak{g} \cong sl(n)$.  It allows us to
establish a one-to-one correspondence between the element $F(x)
\in \mathcal{ M} $ and its expansion coefficients. Indeed, from
eqs. (\ref{3.5'}) - (\ref{3.5'''}) we  can easily prove the
following:
\begin{Prop}\label{pro:V.1}
The function $F(x)\equiv 0 $ if and only if all its expansion
coefficients vanish, i.e.:
\begin{subequations}\label{eqB:pr1}
\begin{eqnarray}\label{eqB:pr1a}
\begin{split}
\gamma^+ _{F;ir}(\lambda)&=\gamma^- _{F;ri}(\lambda)=0, \qquad i <
r;\qquad Z_{F;j}^+(x)&= Z_{F;j}^-(x)=0 ;
\end{split}\\
\label{eqB:pr1b}
\begin{split}
\gamma^+ _{F;ri}(\lambda)&=\gamma^- _{F;ir}(\lambda)=0, \qquad i <
r;\qquad \tilde{Z}_{F;j}^+(x)&= \tilde{Z}_{F;j}^-(x)=0;
\end{split}
\end{eqnarray}
\end{subequations}
where $j=1,\dots, N$.
\end{Prop}

{\sl Proof.}\, To show that from $F(x)\equiv 0 $ there follows
(\ref{eqB:pr1a}) we insert $F(x) \equiv 0 $ into the r.h. sides of
the inversion formulae (\ref{3.5'})-(\ref{3.5'''}) getting
(\ref{eqB:pr1}). The fact that from (\ref{eqB:pr1a}) there follows
$F(x)\equiv 0 $ is obtained by inserting (\ref{eqB:pr1a}) into the
r.h. side of (\ref{3.5'})-(\ref{3.5'''}). The equivalence of
$F(x)\equiv 0 $ to (\ref{eqB:pr1b}) is proved analogously.

\subsection{Expansions of $Q(x) $.}\label{ssec:4.1}

Here we evaluate the expansion coefficients for $F(x) \equiv Q(x)
$. As the reader have guessed already, their evaluation will be
based on the Wronskian relations (\ref{eq:21.2}),
(\ref{eqB:wr.1ab}) which we derived above. From them we have:
\begin{eqnarray}\label{eq:49.1a}
\biglb \e^+_{li}(x,\lambda), Q(x)\bigrb = \ri \d^-_{il}(\lambda),
\qquad
\biglb \e^-_{il}(x,\lambda), Q(x)\bigrb = \ri \d^+_{li}(\lambda),\\
\label{eq:49.1b} \biglb \e^+_{il}(x,\lambda), Q(x)\bigrb = \ri
\b^+_{li}(\lambda), \qquad \biglb \e^-_{li}(x,\lambda), Q(x)\bigrb
= \ri \b^-_{il}(\lambda),
\end{eqnarray}

Skipping the calculational details  we get the following expansion
of $Q(x) $ over the systems $ \{\bPhi ^\pm \}$ and $ \{\bPsi ^\pm
\}$:
\begin{eqnarray}\label{eq:49.4}
\begin{split}
Q(x) &= {\ri\over \pi } \int_{-\infty }^{\infty } \rd\lambda
\sum_{i<r} \left( \tau^+_{ir}(\lambda ) \bPhi_{ir} ^+(x, \lambda )
-\tau_{ri}^-(\lambda )  \bPhi_{ri} ^-(x, \lambda ) \right) \\
&+ 2\sum_{k=1}^{N} \sum_{i<r} \left(\tau^+_{ir;j} \bPhi_{ir;j}
^+(x) + \tau^-_{ri;j} \bPhi_{ri;j} ^-(x)\right),
\end{split}\\
\label{eq:49.5}
\begin{split}
Q(x) &=- {\ri \over \pi } \int_{-\infty }^{\infty } \rd\lambda
\sum_{i<r} \left( \rho^+_{ri}(\lambda ) \bPsi_{ri} ^+(x, \lambda )
-\rho_{ir}^-(\lambda )  \bPsi_{ir} ^-(x, \lambda ) \right) \\
&- 2\sum_{k=1}^{N} \sum_{i<r} \left(\rho^+_{ri;j} \bPsi_{ri;j}
^+(x) + \rho^-_{ir;j} \bPsi_{ir;j} ^-(x)\right),
\end{split}
\end{eqnarray}

\subsection{Expansions of $\ad_J^{-1} \delta Q(x) $.}\label{ssec:4.2}

Here we evaluate the expansion coefficients for $F(x)
\equiv\ad_J^{-1}\delta Q(x) $.  Their evaluation is based on the
Wronskian relations (\ref{eqB:wr.19}), (\ref{eqB:wr.20}). Now we
have:
\begin{eqnarray}\label{eq:50.1a}
\begin{split}
\biglb \e^+_{li}(x,\lambda), \ad_J^{-1}\delta Q(x)\bigrb &=
{\ri\over 2} ((\delta\tau^+) \c^+(\lambda))_{il}, \\ \biglb
\e^-_{il}(x,\lambda), \ad_J^{-1}\delta Q(x)\bigrb &= -{\ri\over 2}
((\delta\tau^-) \c^-(\lambda))_{li},
\end{split}\\ \label{eq:50.1b}
\begin{split}
 \biglb \e^+_{il}(x,\lambda), \ad_J^{-1}\delta
Q(x)\bigrb &= -{\ri\over 2} ((\delta\rho^+) \a^+(\lambda))_{li}, \\
\biglb\e^-_{li}(x,\lambda), \ad_J^{-1}\delta Q(x)\bigrb &=
{\ri\over 2} ((\delta\rho^-) \a^-(\lambda))_{il},
\end{split}
\end{eqnarray}

Skipping the calculational details  we get the following expansion
of $\ad_J^{-1}\delta Q(x) $ over the systems $ \{\bPhi ^\pm \}$
and $ \{\bPsi ^\pm \}$:
\begin{eqnarray}\label{eq:50.6}
\begin{split}
\ad_J^{-1}\delta Q(x) &= {\ri\over 2\pi } \int_{-\infty }^{\infty
} \rd\lambda \sum_{i<r} \left( \delta\tau^+_{ir}(\lambda )
\bPhi_{ir} ^+(x, \lambda )
+ \delta \tau_{ri}^-(\lambda )  \bPhi_{ri} ^-(x, \lambda ) \right) \\
&+ \sum_{k=1}^{N} \sum_{i<r} \left(\delta' W^+_{ir;j}(x) -
\delta'W^-_{ri;j}(x) \right), \end{split}\\
\label{eq:51.4}
\begin{split}
\ad_J^{-1}\delta Q(x) &= {\ri  \over 2\pi } \int_{-\infty
}^{\infty } \rd\lambda \sum_{i<r} \left(\delta \rho^+_{ri}(\lambda
) \bPsi_{ri} ^+(x, \lambda )
+ \delta\rho_{ir}^-(\lambda )  \bPsi_{ir} ^-(x, \lambda ) \right) \\
&+\sum_{k=1}^{N} \sum_{i<r} \left(\delta'\tilde{W}^+_{ir;j}(x) -
\delta'\tilde{W}^-_{ri;j}(x)\right),
\end{split}
\end{eqnarray}
where
\begin{eqnarray}\label{eq:50.6W}
\delta' W^\pm_{ab;j}(x)=\delta\lambda_j^\pm \tau^\pm_{ab;j}
 \dot{\bPhi}_{ab;j} ^\pm(x) + \delta \tau^\pm_{ab;j}
\bPhi_{ab;j} ^\pm(x), \\
\delta'\tilde{W}^\pm_{ab;j}(x)=\delta\lambda_j^\pm \rho^\pm_{ab;j}
\dot{\bPsi}_{ab;j} ^\pm (x) + \delta \rho^\pm_{ab;j} \bPsi_{ab;j}
^\pm(x)
\end{eqnarray}

The expansions (\ref{eq:49.4}), (\ref{eq:49.5}) combined with
proposition \ref{pro:V.1} is another way to establish the
one-to-one correspondence between $Q(x) $ and each of the minimal
sets of scattering data $\mathcal{T}_1 $ and $\mathcal{T}_2 $
(\ref{eq:T_12}). Likewise the expansions (\ref{eq:50.6}),
(\ref{eq:51.4}) and proposition \ref{pro:V.1} establish the
one-to-one correspondence between the variation of the potential
$\delta Q(x) $ and the variations of the scattering data $\delta
\mathcal{T}_1 $ and $\delta \mathcal{T}_2 $.

\subsection{The generating operators}\label{ssec:Lambda}

To complete the analogy between the standard Fourier transform and
the expansions over the `squared solutions' we need the analogs of
the operator $D_0=-\ri \rd/\rd x $. The operator $D_0 $ is the one
for which $\re^{\ri\lambda x} $ is an eigenfunction:  $D_0
\re^{\ri\lambda x}=\lambda \re^{\ri\lambda x} $. Therefore it is
natural to introduce the generating operators $\Lambda _\pm $
through:
\begin{equation}\label{eq:**0}
\begin{split}
(\Lambda _+-\lambda )\bPsi_{ri}^{+} (x,\lambda ) = 0, \qquad
(\Lambda _+-\lambda )\bPsi_{ir}^{-} (x,\lambda ) = 0, \\
(\Lambda _--\lambda )\bPhi_{ir}^{+} (x,\lambda ) = 0, \qquad
(\Lambda _--\lambda )\bPhi_{ri}^{-} (x,\lambda ) = 0.
\end{split}
\end{equation}
Their derivation starts by introducing the splitting:
\begin{equation}\label{eq:**1}
e_{ab}^{\pm}(x,\lambda ) = e_{ab}^{{\rm d},\pm}(x,\lambda ) +
\e_{ab}^{\pm}(x,\lambda ), \qquad e_{ab}^{{\rm d},\pm}(x,\lambda )
= (\openone -P_{0J}) e_{ab}^{\pm}(x,\lambda ),
\end{equation}
into the equation (\ref{eq:e-ab1}). Then eq. (\ref{eq:e-ab1})
splits into:
\begin{eqnarray}\label{eq:**2a}
\ri {\rd e_{ab}^{{\rm d},\pm}\over \rd x}+
[Q(x),\e_{ab}^{\pm}(x,\lambda
)] &=&0,\\
\label{eq:**2b} \ri {\rd \e_{ab}^{\pm}\over \rd x} + [Q(x),
e_{ab}^{{\rm d},\pm}(x,\lambda )] &=& \lambda [J,
\e_{ab}^{\pm}(x,\lambda )],
\end{eqnarray}
Eq. (\ref{eq:**2a}) can be integrated formally with the result
\begin{eqnarray}\label{eq:**3}
e_{ab}^{{\rm d},\pm}(x,\lambda ) = C_{ab;\epsilon }^{{\rm
d},\pm}(\lambda ) + \ri \int_{\epsilon \infty }^{x} \rd y\,
[Q(y),\e_{ab}^{\pm}(y,\lambda)], \\
\label{eq:**3'} C_{ab;\epsilon }^{{\rm d},\pm}(\lambda ) =
\lim_{y\to\epsilon \infty } e_{ab}^{{\rm d},\pm}(y,\lambda ),
\qquad \epsilon =\pm 1.
\end{eqnarray}
Next insert (\ref{eq:**3}) into (\ref{eq:**2b}) and act on both
sides by $\ad_{J}^{-1} $. This gives us:
\begin{eqnarray}\label{eq:**4}
(\Lambda _\pm -\lambda )\e_{ab}^{\pm}(x,\lambda ) =  \ri
[C_{ab;\epsilon }^{{\rm d},\pm}(\lambda ),\ad_{J}^{-1}Q(x) ],
\end{eqnarray}
where the generating operators $\Lambda _\pm $ are given by:
\begin{equation}\label{eq:**6}
\Lambda _\pm X(x) \equiv \ad_{J}^{-1} \left( \ri {\rd X \over \rd
x} + \ri \left[ Q(x), \int_{\pm\infty }^{x} \rd y\, [Q(y),
X(y)]\right] \right).
\end{equation}
Thus $\e_{ab}^{\pm}(x,\lambda ) $ will be an eigenfunction of
$\Lambda _+ $ or $\Lambda _- $ if only if $C_{ab}^{{\rm
d},\pm}(y,\lambda )=0 $. Evaluating the limits of (\ref{eq:**3'})
for all combinations of indices $a,b $ we find ($i<r $):
\begin{eqnarray}\label{eq:**7+}
\begin{split}
(\Lambda _+ -\lambda ) \bPsi_{ri}^{+}(x,\lambda )=0, \qquad
(\Lambda _+ -\lambda ) \bPsi_{ir}^{-}(x,\lambda )=0,  \\
(\Lambda _+ -\lambda_j^+ ) \bPsi_{ri;j}^{+}(x)=0, \qquad (\Lambda
_+ -\lambda_j^- ) \bPsi_{ir;j}^{-}(x)=0,
\end{split}\\
\label{eq:**7-}
\begin{split}
(\Lambda _- -\lambda ) \bPhi_{ir}^{+}(x,\lambda )=0, \qquad
(\Lambda _- -\lambda ) \bPhi_{ri}^{-}(x,\lambda )=0,  \\
(\Lambda _- -\lambda_j^+ ) \bPhi_{ir;j}^{+}(x)=0, \qquad (\Lambda
_- -\lambda_j^- ) \bPhi_{ri;j}^{-}(x)=0,
\end{split}
\end{eqnarray}
The rest of the squared solutions are not eigenfunctions of
neither $\Lambda _+ $ nor $\Lambda _-$:
\begin{eqnarray}\label{eq:**8}
\begin{split}
(\Lambda _+ -\lambda_j^+ )
\dot{\bPsi}_{ri;j}^{+}(x)=\bPsi_{ri;j}^{+}(x), \qquad (\Lambda _+
-\lambda_j^- ) \dot{\bPsi}_{ir;j}^{-}(x)=
\bPsi_{ir;j}^{-}(x), \\
(\Lambda _- -\lambda_j^+ )
\dot{\bPhi}_{ir;j}^{+}(x)=\bPhi_{ir;j}^{+}(x), \qquad (\Lambda _-
-\lambda_j^- ) \dot{\bPhi}_{ri;j}^{-}(x)= \bPhi_{ri;j}^{-}(x),
\end{split}
\end{eqnarray}
i.e., $\dot{\bPsi}_{ri;j}^{+}(x) $ and $\dot{\bPhi}_{ir;j}^{+}(x)
$ are adjoint eigenfunctions of $\Lambda _+ $ and $\Lambda _- $.
This means that $\lambda _j^\pm $, $j=1,\dots, N $ are also the
discrete eigenvalues of $\Lambda _\pm $ but the corresponding
eigenspaces of $\Lambda _\pm $ have double the dimensions of the
ones of $L $; now they are spanned by both $\bPsi_{ab;j}^{\pm}(x)
$ and $\dot{\bPsi}_{ab;j}^{\pm}(x) $. Thus the sets $\{\Psi \} $
and $\{\Phi \} $ are the complete sets of eigen- and adjoint
functions of $\Lambda _+ $ and $\Lambda _- $.

\section{Fundamental Properties of the MNLS-type equations}
In this Section we describe the fundamental properties of the
NLEE.

\subsection{The class of the MNLS-type equations}\label{ssec:5.1}

Let us insert variations of the form (\ref{eqB:wr.23}) into the
expansions (\ref{eq:50.6}), (\ref{eq:51.4}). This results in:
\begin{eqnarray}\label{eq:5.1.1}
\begin{split}
\ad_J^{-1}{d Q\over dt} &= {\ri\over 2\pi } \int_{-\infty
}^{\infty } \rd\lambda \sum_{i<r} \left( {\rd\tau^+_{ir} \over \rd
t} \bPhi_{ir} ^+(x, \lambda )
+ {\rd \tau_{ri}^-\over \rd t} \bPhi_{ri} ^-(x, \lambda ) \right) \\
&+ \sum_{k=1}^{N} \sum_{i<r} \left( W^{\prime,+}_{ir;j}(x,t) -
W^{\prime,-}_{ri;j}(x,t) \right), \end{split}\\
\label{eq:5.1.2}
\begin{split}
\ad_J^{-1}{d Q\over dt} &= {\ri\over 2\pi } \int_{-\infty
}^{\infty } \rd\lambda \sum_{i<r} \left({\rd\rho^+_{ri}\over \rd
t} \bPsi_{ri} ^+(x, \lambda )
+ {\rd\rho_{ir}^-\over \rd t}  \bPsi_{ir} ^-(x, \lambda ) \right) \\
&+\sum_{k=1}^{N} \sum_{i<r} \left(\tilde{W}^{\prime,+}_{ir;j}
(x,t) -\tilde{W}^{\prime,-}_{ri;j}(x,t)\right),
\end{split}
\end{eqnarray}
where
\begin{eqnarray}\label{eq:5.1.1'}
W^{\prime,\pm}_{ab;j}(x,t)={\rd\lambda_j^\pm\over \rd t}
\tau^\pm_{ab;j} \dot{\bPhi}_{ab;j} ^\pm(x) + {\rd\tau^\pm_{ab;j}
\over \rd t}
\bPhi_{ab;j} ^\pm(x), \\
\tilde{W}^{\prime,\pm}_{ab;j}(x)={\rd\lambda_j^\pm\over \rd t}
\rho^\pm_{ab;j} \dot{\bPsi}_{ab;j} ^\pm (x) +
{\rd\rho^\pm_{ab;j}\over \rd t} \bPsi_{ab;j} ^\pm(x)
\end{eqnarray}
Next from (\ref{eq:**7+}), (\ref{eq:**7-}) there follows that:
\begin{eqnarray}\label{eq:5.1.3}
&& \begin{split} && (f_0(\Lambda _+) -f_0(\lambda ))
\bPsi_{ri}^{+}(x,\lambda )=0, \qquad (f_0(\Lambda _+) -f_0(\lambda
)) \bPsi_{ir}^{-}(x,\lambda )=0,
\\[10pt]
&& (f_0(\Lambda _+) -f_0(\lambda_j^+) ) \bPsi_{ri;j}^{+}(x)=0,
\qquad (f_0(\Lambda _+) -f_0(\lambda_j^-) ) \bPsi_{ir;j}^{-}(x)=0,
\end{split}\\
&& \begin{split}\label{eq:5.1.3'} && (f_0(\Lambda _+)
-f_0(\lambda_j^+) ) \dot{\bPsi}_{ri;j}^{+}(x)= \dot{f}(\lambda
_j^+) \bPsi_{ri;j}^{+}(x),
\\
&& (f_0(\Lambda _+) -f_0(\lambda_j^-) ) \dot{\bPsi}_{ir;j}^{-}(x)=
\dot{f}(\lambda _j^-)\bPsi_{ir;j}^{-}(x),
\end{split}
\end{eqnarray}
and similar relations between $\Lambda _- $ and its
eigenfunctions. Combining them with the expansions
(\ref{eq:49.4}), (\ref{eq:49.5}) we get:
\begin{eqnarray}\label{eq:5.1.4}
\begin{split}
f_0(\Lambda _-)Q(x) &= {\ri\over \pi } \int_{-\infty }^{\infty }
\rd\lambda f_0(\lambda )\sum_{i<r} \left( \tau^+_{ir}(\lambda )
\bPhi_{ir} ^+(x,
\lambda ) -\tau_{ri}^-(\lambda )  \bPhi_{ri} ^-(x, \lambda ) \right) \\
&+ 2\sum_{k=1}^{N} \sum_{i<r} \left(f_0(\lambda _j^+)\tau^+_{ir;j}
\bPhi_{ir;j} ^+(x) + f_0(\lambda _j^-)\tau^-_{ri;j} \bPhi_{ri;j}
^-(x)\right),
\end{split}\\
\label{eq:49.5f}
\begin{split}
f_0(\Lambda _+)Q(x) &=- {\ri\over \pi } \int_{-\infty }^{\infty }
\rd\lambda \sum_{i<r} f_0(\lambda )\left( \rho^+_{ri}(\lambda )
\bPsi_{ri} ^+(x, \lambda ) -\rho_{ir}^-(\lambda )  \bPsi_{ir}
^-(x, \lambda ) \right)
\\ &- 2\sum_{k=1}^{N} \sum_{i<r} \left(f_0(\lambda _j^+)\rho^+_{ri;j}
\bPsi_{ri;j} ^+(x) + f_0(\lambda _j^-)\rho^-_{ir;j} \bPsi_{ir;j}
^-(x)\right),
\end{split}
\end{eqnarray}

Now we can prove that the principal series of MNLS-type equations
has the form:
\begin{equation}\label{eq:5.1.5}
\ri \ad_{J}^{-1} {\rd Q  \over \rd t } + f_0(\Lambda )Q(x,t)=0,
\end{equation}
where $\Lambda  $ can be either $\Lambda _+ $ or $\Lambda _- $ and
$f_0(\lambda ) $ determines the dispersion law of the
corresponding NLEE.

\begin{theorem}\label{t1}
The NLEE (\ref{eq:5.1.5}) are equivalent to each of the following
evolution equations for the scattering data of $L$:
\begin{gather}
\begin{split}\label{dsa}
\ri {\rd\rho ^\pm \over \rd t} \mp 2f_0(\lambda)\rho ^\pm = 0,
\qquad {\rd\lambda _j^\pm \over \rd t} =0, \qquad \ri
{\rd\rho_{;j} ^\pm \over \rd t} \mp 2f_0(\lambda _j^\pm)\rho_{;j}
^\pm =0,
\end{split}\\
\begin{split} \label{dsb}
\ri {d\tau ^\pm \over \rd t} \pm 2f_0(\lambda)\tau ^\pm = 0,
\qquad {\rd\lambda _j^\pm \over \rd t} =0, \qquad i {\rd\tau_{;j}
^\pm \over \rd t} \pm 2f_0(\lambda _j^\pm)\tau_{;j} ^\pm =0,
\end{split}
\end{gather}
\end{theorem}

\begin{proof}
Insert the expansions (\ref{eq:5.1.2}) and (\ref{eq:49.5f}) into
the left hand side of the NLEE (\ref{eq:5.1.5}) and use
proposition \ref{pro:V.1}. This immediately proves the equivalence
of the NLEE to the linear equations (\ref{dsa}). Analogously, from
the expansions (\ref{eq:5.1.1}), (\ref{eq:5.1.4}) and proposition
\ref{pro:V.1} one proves the equivalence between the NLEE and
(\ref{dsb}).
\end{proof}

\subsection{Integrals of motion -- principal
series.}\label{ssec:Integr}

Dealing with the block Zakharov-Shabat system we can distinguish
two types of NLEE and, consequently, two types of series of
conservation laws that provide the Hamiltonians of the NLEE.

The MNLS equations have maximally  degenerated dispersion law and
belong to the principle series of NLEE.  Their Hamiltonians have
local densities, i.e., their densities depend on $Q(x,t) $ and its
$x $-derivatives. At the same time they have maximal number of
generating functionals of conservation laws: the whole blocks
$\a^\pm(\lambda ) $ and $\c^\pm(\lambda ) $. However not all of
these functionals are in involution, see Subsection \ref{ssec:5.3}
below.

The NLEE characterized by  generic dispersion laws have as a rule
non-local Hamiltonian densities and a minimal possible number of
generating functionals of integrals of motion. These are provided
by the invariants (the eigenvalues) of $\a^\pm(\lambda ) $ and
$\c^\pm(\lambda ) $. Special combinations of these as in eq.
(\ref{eq:2x2-5'}) produce the Hamiltonians of the MNLS type
equations. In this Subsection we will concentrate on those.

In Subsection \ref{ssec:2x} we showed how these functionals
$A^+(\lambda ) $ and $C^+(\lambda ) $ can be expressed through the
minimal sets of scattering data, see eq. (\ref{eq:2x2-6}). Our aim
here is to derive a recurrent procedure which would allow one to
express their coefficients in the asymptotic expansions:
\begin{equation}\label{eq:AC-as}
A^+(\lambda ) = \sum_{k=1}^{\infty } I_k \lambda ^{-k}, \qquad
C^+(\lambda ) = \sum_{k=1}^{\infty } J_k \lambda ^{-k},
\end{equation}
in terms of $Q(x,t) $. To this end we make use  of a Wronskian
type relation:
\begin{equation}\label{eq:J2.1}
(\openone -P_{0J})
\left. \hat{\chi }^+(x,\lambda )\left( \dot{\chi }^+ +\ri xJ\chi
^+\right) \right|_{x=-\infty }^{\infty } = \left(
\begin{array}{cc} \hat{\a}^+\dot{\a}^+  & 0 \\ 0 &
-\hat{\c}^+\dot{\c}^+ \end{array} \right),
\end{equation}
valid for $\lambda \in\bbbc_+ $.  Here and below by `dot' we
denote derivative with respect to $\lambda  $. In what follows we
will need also the standard formula:
\begin{equation}\label{eq:stand}
\tr \ln \a^+(\lambda ) = \ln \det \a^+(\lambda ) ,
\end{equation}
and its consequence
\begin{equation}\label{eq:J2.10}
{\rd \over \rd\lambda }\tr \ln \a^+(\lambda ) = \ln \det
\hat{\a}^+(\lambda ) \dot{\a}^+(\lambda ).
\end{equation}

Next we express $\chi ^+(x,\lambda ) $ in the form:
\begin{equation}\label{eq:J5.10}
\chi ^+(x,\lambda ) = (\openone +W(x,\lambda )) Z(x,\lambda )
\re^{-\ri\lambda Jx},
\end{equation}
where
\begin{equation}\label{eq:J5.11}
W(x,\lambda ) = \left(\begin{array}{cc} 0 & W^{(1)}(x,\lambda ) \\
W^{(2)}(x,\lambda ) & 0 \end{array} \right), \quad
Z(x,\lambda ) = \left(\begin{array}{cc} Z^{(1)}(x,\lambda ) & 0 \\
0 & Z^{(2)}(x,\lambda )  \end{array} \right),
\end{equation}

Inserting (\ref{eq:J5.10}) into (\ref{eq:4.1}) and separating the
block-diagonal and block-off-diagonal parts we get for
$W(x,\lambda ) $ and $Z(x,\lambda ) $ the following system:
\begin{eqnarray}\label{eq:J5.1a}
\ri {\rd W \over \rd x } + Q(x,t) - WQW(x,t,\lambda ) = \lambda
[J,W(x,t,\lambda )] , \\ \label{eq:J5.1b} \ri {\rd Z  \over \rd x
} \hat{Z}(x,t,\lambda ) + Q(x,t)W(x,t,\lambda )=0.
\end{eqnarray}
Combining eqs. (\ref{eq:J5.10}) and (\ref{eq:J2.1}) we obtain:
\begin{equation}\label{eq:J6.1}
\left. \hat{Z}(x,t,\lambda ) \dot{Z} \right|_{x=-\infty }^{\infty
} = \left( \begin{array}{cc} \hat{\a}^+(\lambda )
\dot{\a}^+(\lambda ) & 0
\\
0 & -\hat{\c}^+(\lambda ) \dot{\c}^+(\lambda ) \end{array}\right),
\end{equation}
which leads, in view of (\ref{eq:J2.10}) and (\ref{eq:J5.1b}) to
\begin{equation}\label{eq:J6.2}
\left. \hat{Z}(x,t,\lambda ) \dot{Z} \right|_{x=-\infty }^{\infty
} = \ri \int_{-\infty }^{\infty } \rd x \hat{Z}(x,t,\lambda )
Q(x,t) \dot{W}(x,t,\lambda ) Z(x,t,\lambda ).
\end{equation}
If we multiply both sides of (\ref{eq:J6.2}) by
$\left(\begin{array}{cc} \openone &0 \\ 0 & 0 \end{array}\right) $
and  $\left(\begin{array}{cc} 0 &0 \\ 0 & \openone
\end{array}\right) $ take the trace and integrate over $\lambda  $
we get:
\begin{equation}\label{eq:J6.3}
\begin{split}
A^+(\lambda ) \equiv \ln \det \a^+(\lambda ) = \ri \int_{-\infty
}^{\infty } \rd x\, \tr \left( Q(x,t) W(x,t,\lambda )
\left(\begin{array}{cc}
\openone & 0 \\ 0 & 0 \end{array}\right)\right), \\
C^+(\lambda ) \equiv \ln \det \c^+(\lambda ) = -\ri \int_{-\infty
}^{\infty } \rd x\, \tr \left( Q(x,t) W(x,t,\lambda )
\left(\begin{array}{cc} 0 & 0 \\ 0 & \openone
\end{array}\right)\right).
\end{split} \end{equation}
Using the asymptotic expansions of $A^+(\lambda ) $, $C^+(\lambda
) $ (see eq. (\ref{eq:AC-as})) and $W(x,t,\lambda ) $:
\begin{equation}\label{eq:W-as}
W(x,t,\lambda ) = \sum_{k=1}^{\infty } W_k(x,t)\lambda ^{-k},
\end{equation}
we arrive at the following expressions for the integrals of motion
$I_k $ and $J_k $:
\begin{equation}\label{eq:IJ-k} \begin{split}
I_k = \ri \int_{-\infty }^{\infty } \rd x\, \tr \left( Q(x,t)
W_k(x,t) \left(\begin{array}{cc} \openone  & 0 \\ 0 & 0
\end{array}\right)\right),
\\
J_k = -\ri \int_{-\infty }^{\infty } \rd x\, \tr \left( Q(x,t)
W_k(x,t) \left(\begin{array}{cc} 0& 0 \\ 0 & \openone
\end{array}\right)\right).
\end{split}
\end{equation}

The last step in these considerations consists in deriving
recurrent relations for calculating $W_k(x,t) $ in terms of
$Q(x,t) $ and its $x $-derivatives. This is done by inserting
(\ref{eq:W-as}) into the equation (\ref{eq:J5.1a}) with the
result:
\begin{equation}\label{eq:W-rec}
\begin{split}
W_1(x,t) &= {1  \over 4 } [J,Q(x,t)], \\
W_{k+1}(x,t) &= {1  \over 4 } \left[ J,\; \ri {\rd W_k \over \rd x
} - \sum_{p+s=k}^{} W_p(x,t) Q(x,t) W_s(x,t) \right].
\end{split}
\end{equation}
In particular we have:
\begin{equation}\label{eq:W2-3}
W_2(x,t) = {\ri \over 4 } Q_x, \qquad W_3(x,t) = {1\over 16 }
\left[ J, Q_{xx} + Q^3 \right],
\end{equation}
which leads to:
\begin{eqnarray}\label{eq:IJ1-3}
I_1 &=& {\ri\over 4 } \int_{-\infty }^{\infty } \rd x\, \tr (qr),
\qquad J_1 = -{\ri \over 4 } \int_{-\infty }^{\infty } \rd x\, \tr
(rq),
\nonumber\\
I_2 &=& -{1 \over 4 } \int_{-\infty }^{\infty } \rd x\, \tr
(qr_x), \qquad
J_2 = {1  \over 4 } \int_{-\infty }^{\infty } \rd x\, \tr (rq_x), \\
I_3 &=& -{\ri \over 8 } \int_{-\infty }^{\infty } \rd x\, \tr
(qr_{xx}+qrqr ), \qquad J_3 = -{\ri \over 8 } \int_{-\infty
}^{\infty } \rd x\, \tr (rq_{xx} +rqrq).  \nonumber
\end{eqnarray}

\subsection{Hamiltonian properties of the MNLS eqs.}\label{ssec:Ham}

Let us briefly outline the Hamiltonian properties of the NLEE
(\ref{eq:5.1.5}). Obviously the MNLS describes an infinite
dimensional Hamiltonian system with Hamiltonian:
\begin{equation}\label{eq:H-MNLS}
H_{\rm MNLS} = {1 \over 2}\int_{-\infty }^{\infty } \rd y\, \tr
\left( QQ_{xx} + Q^4(x,t) \right) = 4\ri (I_3+J_3),
\end{equation}
and Poisson brackets:
\begin{equation}\label{eq:PB}
\{\q_{ks}(y,t) , \r_{ri}(x,t)  \} = \ri \delta _{ik}\delta _{rs}
\delta (x-y),
\end{equation}
or, equivalently, by the canonical symplectic form:
\begin{eqnarray}\label{eq:Ome0}
\Omega _0&=& \ri \int_{-\infty }^{\infty }\rd x\, \tr \left(\delta
\r(x)
\wedge \delta \q(x) \right)\nonumber\\
&=& {1 \over \ri} \int_{-\infty }^{\infty } \rd x\, \tr \left(
\ad_{J}^{-1} \delta Q(x) \wedge [ J , \ad_{J}^{-1} \delta Q(x)
\right).
\end{eqnarray}

The second expression is preferable to us because it makes obvious
the interpretation of $\delta Q(x,t) $ as local coordinate on the
co-adjoint orbit passing through $J $. It is also expressed
through the skew-scalar product by:
\begin{equation}\label{eq:Ome0s}
\Omega _0 = {1 \over \ri} \biglb \ad_{J}^{-1} \delta Q \wedgecomma
\ad_{J}^{-1} \delta Q \bigrb.
\end{equation}

It can be evaluated in terms of the scattering data variations. To
do this we insert the expansion of $\ad_{J}^{-1} \delta Q $ into
$\Omega _0 $ and then use again the equations (\ref{eq:50.1a}),
(\ref{eq:50.1b}). After some calculations we get:
\begin{eqnarray}\label{eq:5.2.5}
\Omega _0 &=& {1  \over 2\pi \ri} \int_{-\infty }^{\infty }
\rd\lambda \, \left( \Omega _{0}^{+}(\lambda ) -\Omega
_{0}^{-}(\lambda ) \right) -\ri \sum_{j=1}^{N} \left( \Omega
_{0,j}^{+}+
\Omega_{0;j}^{-}\right) , \\
\Omega _{0}^{\pm}(\lambda )&=& {1  \over 2 } \tr \left( \delta
\tau^\pm \c^\pm \wedge \delta \rho ^\pm \a^\pm(\lambda )\right),
\qquad \Omega _{0,j}^{\pm} = \Res_{\lambda =\lambda _j^\pm} \Omega
_{0}^{\pm} (\lambda ),
\end{eqnarray}
Here we skip the explicit expressions for $\Omega _{0,j}^{\pm} $
which are rather involved. From (\ref{eq:5.2.5}) it is not even
obvious that $\Omega _0$ is closed.

The Hamiltonian formulation of the MNLS eq. with $\Omega _0 $ and
$H_0 $ is just one member of the hierarchy of Hamiltonian
formulations of MNLS provided by:
\begin{eqnarray}\label{eq:5.2.6}
\Omega _k &=& {1 \over \ri}\biglb \ad_{J}^{-1} \delta Q
\wedgecomma \Lambda ^k \ad_{J}^{-1} \delta Q \bigrb , \qquad
\Lambda ={1 \over 2 }
(\Lambda_+ +\Lambda _-), \\
H_k &=& 4\ri ( I_{k+3} + J_{k+3}).
\end{eqnarray}
We can also calculate $\Omega _k $ in terms of the scattering data
variations. Doing this we will need also eqs. (\ref{eq:5.1.3}),
(\ref{eq:5.1.3'}). The answer is
\begin{eqnarray}\label{eq:5.2.6a}
\Omega _k &=&  {1  \over 2\pi \ri} \int_{-\infty }^{\infty }
\rd\lambda \, \lambda ^k \left( \Omega _{0}^{+}(\lambda ) -\Omega
_{0}^{-}(\lambda ) \right) -\ri \sum_{j=1}^{N} \left( \Omega
_{k,j}^{+}+
\Omega_{k;j}^{-}\right), \\
\Omega _{k,j}^{\pm}& = & \Res_{\lambda =\lambda _j^\pm}\lambda ^k
\Omega _{0}^{\pm} (\lambda ).
\end{eqnarray}

This allows one to prove that if we are able to cast $\Omega_{0} $
in canonical form  then all $\Omega _k $ will also be cast in
canonical form and will be pair-wise equivalent.

\subsection{The classical $R$-matrix and the NLEE of MNLS type}
\label{ssec:5.3}

     One of the definitions of the classical $R$-matrix is based on
the Lax representation for the corresponding NLEE. We will start
from this definition, but before to state it will introduce the
following notation:
\begin{equation}\label{eq:L5.1}
\left\{U(x,\lambda ) \otimescomma U(y,\mu )\right\}  ,
\end{equation}
which is an abbreviated record for the Poisson bracket between all
matrix elements of $U(x,\lambda )$ and $U(y,\mu )$
\begin{equation}\label{eq:L5.2}
\left\{ U(x,\lambda ) \otimescomma U(y,\mu )\right\}_{ik,lm} =
\left\{U_{ik}  (x,\lambda ), U_{lm}  (y,\mu )\right\} .
\end{equation}
In particular, if $U(x,\lambda )$ is of the form:
\begin{equation}\label{eq:L5.3}
U(x,\lambda ) = Q(x,t) - \lambda J  ,\qquad Q(x,t) = \sum_{i<r}^{}
(q_{ir}E_{ir}  + p_{ri} E_{ri}) ,
\end{equation}
and the matrix elements of $Q(x,t)$ satisfy (\ref{eq:PB}), then:
\begin{equation}\label{eq:L5.4}
\left\{U(x,\lambda ) \otimescomma U(y,\mu )\right\} = \ri
\sum_{i<r}^{}( E_{ir}\otimes E_{ri} - E_{ri}\otimes E_{ir}) \delta
(x-y) .
\end{equation}

The classical $R$-matrix can be defined through the relation
\cite{FaTa}:
\begin{equation}\label{eq:L5.5}
\left\{U(x,\lambda ) \otimescomma  U(y,\mu )\right\} =
\ri\left[R(\lambda -\mu ),U(x,\lambda )\otimes \openone  +
\openone \otimes U(y,\mu )\right] \delta (x-y).
\end{equation}
which can be understood as a system of $N^2 $ equation for the
$N^2 $ matrix elements of $R(\lambda -\mu )$. However, these
relations must hold identically with respect to $\lambda $ and
$\mu $, i.e., (\ref{eq:L5.5}) is an overdetermined system of
algebraic equations for the matrix elements of $R$. It is far from
obvious whether such $R(\lambda -\mu )$ exists, still less obvious
is that it depends only on the difference $\lambda -\mu $. In
other words far from any choice for $U(x,\lambda )$ and for the
Poisson brackets between its matrix elements allow $R$-matrix
description. Our system  (\ref{eq:L5.5}) allows an  $R$--matrix
given by:
\begin{equation}\label{eq:L5.6}
R(\lambda -\mu ) = - {\ri  \over 2 } { P \over \lambda  - \mu  } ,
\end{equation}
where $P$ is a constant $N^2\times N^2$ matrix:
\begin{equation}\label{eq:L5.7}
P = \sum_{a,b=1}^{N} E_{ab}\otimes E_{ba}  .
\end{equation}

     The matrix $P$ possesses the following special properties:
\begin{equation}\label{eq:L5.8}
P(X \otimes Y) = (Y \otimes X)P  ,\qquad  P^2 \equiv \openone   ,
\end{equation}
i.e., it interchanges the positions of the elements in the direct
tensor product. By using these properties of $P$ we are getting:
\begin{equation}\label{eq:L5.9}
\left[ P, Q(x)\otimes \openone  + \openone \otimes Q(x)\right] = 0
,
\end{equation}
i.e., the r.h.side of (\ref{eq:L5.5}) does not contain $Q(x,t)$.
Besides:
\begin{eqnarray}\label{eq:L5.10}
&& \left[P, \lambda J\otimes \openone  + \mu \openone \otimes
J\right] =
(\lambda  - \mu ) \left[P, J \otimes \openone \right] \nonumber\\
&& = - 2(\lambda  - \mu )\left( \sum_{i<r}^{}( E_{ir}\otimes
E_{ri} - E_{ri}\otimes E_{ir}) \right) .
\end{eqnarray}
where we used the commutation relations between the matrices
$E_{ab} $:
\begin{equation}\label{eq:L5.11}
\left[E_{ab}, E_{cd}\right] = E_{ad}\delta _{bc} -E_{cb}\delta
_{da}.
\end{equation}
The comparison between (\ref{eq:L5.9}), (\ref{eq:L5.10}) and
(\ref{eq:L5.5}) leads us to the result, that $R(\lambda -\mu )$
(\ref{eq:L5.6}) indeed satisfies the definition (\ref{eq:L5.5}).

Let us now show, that the classical $R$--matrix is a very
effective tool for calculating the Poisson brackets between the
matrix elements of $T (\lambda )$. It will be more convenient here
to consider periodic boundary conditions on the interval $[-L,L]
$, i.e. $Q(x-L)=Q(x+L) $ and to use the fundamental solution
$T(x,y,\lambda ) $ defined by:
\begin{equation}\label{eq:eqT}
\ri {\rd T(x,y,\lambda )  \over \rd x } + U(x,\lambda
)T(x,y,\lambda ) =0, \qquad T(x,x,\lambda )=\openone .
\end{equation}

Skipping the details we just formulate the following relation for
the Poisson brackets between the matrix elements of $T(x,y,\lambda
) $:
\begin{equation}\label{eq:L5.12}
\left\{ T(x,y,\lambda ) \otimescomma  T(x,y,\mu )\right\} = \left[
R(\lambda -\mu ), T(x,y,\lambda )\otimes T(x,y,\mu )\right]
\end{equation}

The corresponding monodromy matrix $T_L(\lambda ) $ describes the
transition from $-L $ to $L $ and $T_L(\lambda ) =T(-L,L,\lambda
)$. The Poisson brackets between the matrix elements of
$T_L(\lambda )$ follow directly from eq. (\ref{eq:L5.12}) and are
given by:
\begin{equation}\label{eq:L5.20}
\left\{T_L (\lambda ) \otimescomma T_L(\mu )\right\} =
\left[R(\lambda -\mu ), T_L(\lambda ) \otimes T_L (\mu )\right]  .
\end{equation}

An elementary consequence of this result is the involutivity of
the integrals of motion $I_{L,k}$ from the principal series which
are from the expansions of:
\begin{eqnarray}\label{eq:ln-det}
\ln \det \a_L^+(\lambda ) = \sum_{k=1}^{\infty } I_{L,k}\lambda
^{-k}, \qquad  -\ln \det \c_L^-(\lambda ) = \sum_{k=1}^{\infty }
I_{L,k}\lambda
^{-k}, \\
\ln \det \c_L^+(\lambda ) = \sum_{k=1}^{\infty } J_{L,k}\lambda
^{-k}, \qquad  -\ln \det \a_L^-(\lambda ) = \sum_{k=1}^{\infty }
J_{L,k}\lambda ^{-k},
\end{eqnarray}
An important property of the integrals $I_{L,k} $ and $J_{L,k} $
is their locality, i.e. their densities depend only on $Q $ and
its $x $-derivatives.

The simplest consequence of the  relation (\ref{eq:L5.12}) is the
involutivity of $I_{L,k} $, $J_{L,k} $. Indeed, taking the trace
of both sides of (\ref{eq:L5.12}) shows that $\{ \tr T_L(\lambda
),\tr T_L(\mu )\}=0 $.  We can also multiply both sides of
(\ref{eq:L5.12}) by $C\otimes C$ and then take the trace using eq.
(\ref{eq:L5.9}); this proves:

\begin{equation}\label{eq:L5.23}
\left\{\tr T_L (\lambda )C, \tr T_L (\mu )C\right\} = 0 .
\end{equation}
In particular, for $C = \openone +J$  and $C = \openone -J$  we
get the involutivity of:
\begin{equation}\label{eq:L5.24}
\begin{split}
\left\{ \tr \a^+_L(\lambda ),\tr \a^+_L(\mu )\right\} =0, \qquad
\left\{\tr \a^-_L(\lambda ),\tr \a^-_L(\mu )\right\} = 0 ,\\
\left\{ \tr \c^+_L(\lambda ),\tr \c^+_L(\mu )\right\} =0, \qquad
\left\{\tr \c^-_L(\lambda ),\tr \c^-_L(\mu )\right\} = 0 ,
\end{split}
\end{equation}

Eq. (\ref{eq:L5.12}) was derived for the typical representation
$V^{(1)} $ of $\mathfrak{G}\simeq SU(n+m) $, but it holds true
also for any other finite-dimensional representation of
$\mathfrak{G} $. Let us denote by $V^{(k)}\simeq \wedge^{k}V^{(1)}
$ the $k $-th fundamental representation of $\mathfrak{G} $; then
the element $T_L(\lambda ) $ will be represented in $V^{(k)} $ by
$\wedge^k T_L(\lambda ) $ -- the $k $-th wedge power of
$T_L(\lambda ) $, see \cite{Helg}. In particular, if we consider
eq. (\ref{eq:L5.12}) in the representation $V^{(n)} $ and sandwich
it between the highest and lowest weight vectors in $V^{(n)} $ we
get:
\begin{equation}\label{eq:det-a}
\{ \det \a^+_L(\lambda ), \det \a^+_L(\mu )\} =0, \qquad \{ \det
\c^-_L(\lambda ), \det \c^-_L(\mu )\} =0.
\end{equation}

Likewise considering (\ref{eq:L5.12}) in the representation
$V^{(m)} $ and sandwich it between the highest and lowest weight
vectors in $V^{(m)} $ we get:
\begin{equation}\label{eq:det-am}
\{ \det \a^-_L(\lambda ), \det \a^-_L(\mu )\} =0, \qquad \{ \det
\c^+_L(\lambda ), \det \c^+_L(\mu )\} =0.
\end{equation}
Since eqs. (\ref{eq:det-a}) and (\ref{eq:det-am}) hold true for
all values of $\lambda  $ and $\mu  $ we can insert into them the
expansions (\ref{eq:ln-det}) with the result:
\begin{equation}\label{eq:I_L}
\{ I_{L,k}, I_{L,p}\} =0,\qquad \{ J_{L,k}, J_{L,p}\} =0,\qquad
k,p= 1,2,\dots.
\end{equation}

Somewhat more general analysis along this lines allows one to see
that only the eigenvalues of $\a_L^\pm(\lambda ) $ and
$\c_L^\pm(\lambda ) $ produce integrals of motion in involution.

Taking the limit $L \to \infty $ we are able to transfer these
results also for the case of potentials with zero boundary
conditions. Indeed, let us multiply (\ref{eq:L5.12}) by
$E(y,\lambda ) \otimes E(y,\mu )$ on the right and by $E^{-1}
(x,\lambda ) \otimes E^{-1} (x,\mu )$  on the left, where
$E(x,\lambda)=\exp (-\ri\lambda Jx)$ and take the limit for $x \to
\infty $, $y \to -\infty $. Since:
\begin{equation}\label{eq:L5.26}
\lim_{x\to\pm \infty } {\re^{\ri x(\lambda -\mu )} \over \lambda
- \mu } =\pm \ri \pi  \delta (\lambda  - \mu ),
\end{equation}
we get:
\begin{eqnarray}\label{eq:L5.27}
\begin{split}
&\left\{T (\lambda ) \otimescomma T (\mu )\right\} = R_+ (\lambda
- \mu )T (\lambda ) \otimes T (\mu ) -
T (\lambda ) \otimes T (\mu ) R_- (\lambda -\mu ) , \\
&R_\pm (\lambda -\mu ) \\
 &= - {1\over 2(\lambda  - \mu ) } \left(\sum_{ik} E_{ik} \otimes
E_{ki}+ \sum_{rs} E_{rs} \otimes E_{sr} \right) \pm \ri\pi \delta
(\lambda -\mu )\Pi_{0J},
\end{split}
\end{eqnarray}
where $\Pi_{0J}$ is defined by eq. (\ref{eq:5.23'}).
Analogously we prove that: \\
i) the integrals $I_k=\lim_{L\to\infty }I_{L,k} $ and
$J_p=\lim_{L\to\infty }J_{L,p} $ are in involution, i.e.:
\[ \{I_k, I_p\} =  \{I_k, J_p\} =  \{J_k, J_p\} = 0, \]
for all positive values of $k $ and $p $; ii) only the eigenvalues
of $\a^\pm(\lambda ) $ and $\c^\pm(\lambda ) $ produce integrals
of motion in involution.

\subsection{Generic NLEE}

Now we apply the expansions above to the analysis of the generic
NLEE related to $L$. Each of these NLEE is determined by its
dispersion law:
\begin{equation}\label{dl}
f(\lambda) = \sum_{k=1}^{r} \lambda^k B_k =\left(
\begin{array}{cc} f_1(\lambda ) & 0 \\ 0 & -f_2(\lambda )
\end{array}\right) , \qquad B_k = \left( \begin{array}{cc} B_{k,1}
& 0 \\ 0 &  -B_{k,2} \end{array}\right),
\end{equation}
The analysis of these NLEE is based on the expansions
$[B_k,\ad_J^{-1} Q(x,t)]$ over the systems $\{\bPsi \}$ and
$\{\bPhi \}$. The corresponding expansion coefficients are
obtained by multiplying eq. (\ref{eq:21.2}) on the right by $B_k$
and taking the trace. Then apply $\Lambda^k$ to both sides of the
expansions. This  proves a theorem generalizing theorem \ref{t1}.

\begin{theorem}\label{t4}
The generic NLEE with polynomial dispersion law $f(\lambda)$
(\ref{dl}) are of the form:
\begin{equation}\label{eq:xx-1}
\ri \ad_{J}^{-1} {\rd Q  \over \rd t } + \sum_{k=1}^r \Lambda^k
[B_k, \ad_J^{-1}Q(x,t)]=0,
\end{equation}
and are equivalent to each of the following evolution equations
for the scattering data of $L$:
\begin{gather}\label{dsa-g}
\begin{split}
\ri {\rd\rho ^+\over \rd t} - f_2(\lambda)\rho ^+(\lambda ,t) -
\rho
^+(\lambda ,t) f_1(\lambda ) = 0, \\
\ri {\rd\rho ^-\over \rd t} + f_1(\lambda)\rho ^-(\lambda ,t) +
\rho
^-(\lambda ,t) f_2(\lambda ) = 0, \\
\ri {\rd\rho_j ^+\over \rd t} - f_{2,j}^+\rho_j ^+(t) - \rho_j
^+(t)
f_{1,j}^+ = 0, \qquad f_{1,2;j}^\pm =f_{1,2} (\lambda_j^\pm),  \\
\ri {\rd\rho_j ^-\over \rd t} + f_{1,j}^-\rho _j^-(t) -\rho_j
^-(t) f_{2,j}^-= 0,\qquad {\rd\lambda _j^\pm \over \rd t} =0,
\end{split}\\ \label{dsb-g}
\begin{split}
\ri {\rd\tau ^+\over \rd t} + f_1(\lambda)\tau ^+(\lambda ,t) +
\tau
^+(\lambda ,t) f_2(\lambda ) = 0, \\
\ri {\rd\tau ^-\over \rd t} - f_2(\lambda)\tau ^-(\lambda ,t) -
\tau
^-(\lambda ,t) f_1(\lambda ) = 0, \\
\ri {\rd\tau_j ^+\over \rd t} + f_{1,j}^+\tau_j ^+(t) + \tau_j
^+(t)
f_{2,j}^+ = 0, \qquad f_{1,2;j}^\pm =f_{1,2} (\lambda_j^\pm) \\
\ri {\rd\tau_j ^-\over \rd t} - f_{2,j}^-\tau _j^-(t) -\tau_j
^-(t) f_{1,j}^-= 0, \qquad {\rd\lambda _j^\pm \over \rd t} =0.
\end{split}
\end{gather}
\end{theorem}

As a consequence of theorem \ref{t4} we get that $S^\pm(\lambda)$
and $T^\pm(\lambda)$ satisfy linear set of equations:
\begin{equation}\label{xx-2}
\begin{split}
\ri {\rd S^\pm \over \rd t} +
[f(\lambda),S^\pm(\lambda,t)]=0,\qquad \ri {\rd T^\pm \over \rd t}
+ [f(\lambda),T^\pm(\lambda,t)]=0,
\end{split}
\end{equation}
which can be easily solved. In addition the block-diagonal
matrices $D^\pm(\lambda)$ are not integrals of motion but rather
satisfy:
\begin{equation}\label{dd}
\ri {\rd D^\pm \over \rd t} + \left[ f(\lambda), D^\pm(\lambda,t)
\right] = 0,
\end{equation}
From eq. (\ref{dd}) there follows that for generic $f(\lambda)$
only the invariants of $D^\pm (\lambda,t)$ (or, equivalently, only
the invariants of $\a^\pm (\lambda,t)$ and $\c^\pm (\lambda,t)$)
provide series of integrals of motion in involution. As a
consequence the generic NLEE (\ref{eq:xx-1}) possess soliton
solutions whose velocities may depend on time.  Examples of such NLEE
and the properties of their soliton solutions called boomerons and
trappons have been analyzed by Calogero and Degasperis
\cite{CaDe,CaDe2,CaDe-new,D1}.

\section{Discussion}\label{sec:l5-6}

In order to understand better the idea that the expansions over
the `squared solutions' are generalized Fourier transforms we will
outline the limit to small potentials, i.e. $Q(x) \ll 1$, for
which the Born approximation is adequate. We will see that in this
limit all these expansions turn out to be just the usual Fourier
transforms.

It is known \cite{AKNS} that the corresponding Zakharov-Shabat system may
have discrete eigenvalues only provided $Q(x) $ is large enough, while in
the Born approximation it does not possess discrete spectrum.  The
eigenfunctions of the continuous spectrum are well approximated by the
`plane waves':
\begin{subequations}\label{eqB:3.39}
\begin{eqnarray}\label{eqB:3.39a}
\chi ^+(x,\lambda ) \simeq \chi ^-(x,\lambda ) \simeq
\re^{-\ri\lambda J x} ,
\end{eqnarray}
and the scattering data are provided by the Born approximation:
\begin{eqnarray}\label{eqB:3.39b}
\begin{split}
\rho_{ir} ^+(\lambda ) &= \ri \int_{-\infty }^{\infty } \q_{ir}(x)
\re^{2\ri \lambda x} \, \rd x, \qquad \rho_{ri} ^-(\lambda ) = \ri
\int_{-\infty }^{\infty } \r_{ri}(x)
\re^{ -2\ri\lambda x} \, \rd x, \\
a^+(\lambda ) &\simeq 1, \qquad a^-(\lambda ) \simeq 1,
\end{split}
\end{eqnarray}
\end{subequations}
As a result the ``squared solutions'' are also approximated by
`plane waves':
\begin{eqnarray*}\label{eqB:3.40}
\begin{split}
\bPsi_{ri} ^+(x,\lambda ) \simeq \bPhi _{ri}^-(x,\lambda )\simeq E
_{ir} \re^{2\ri\lambda x},\qquad \bPsi _{ir}^-(x,\lambda ) \simeq
\bPhi_{ri} ^+(x,\lambda ) \simeq E_{ir} \re^{-2\ri\lambda x},
\end{split}
\end{eqnarray*}
The completeness relation (\ref{eq:5.23}) acquires the form:
\begin{eqnarray*}
\delta (x-y) \Pi_{0J}  = {1  \over  \pi} \int_{-\infty }^{\infty }
\rd x\, \sum_{i<r}\left( \re^{-2\ri\lambda (x-y)} E _{ir} \otimes
E _{ri} - \re^{2\ri\lambda (x-y)} E _{ri} \otimes E _{ir} \right),
\end{eqnarray*}
of the usual Fourier transform for matrix-valued functions. The
recursion operators $\Lambda _\pm $ in this limit go into the
purely differentiation operator:
\begin{eqnarray}\label{eqB:3.42}
\Lambda _\pm^{\rm as} \cdot \simeq {\ri  \over 4 } \left[ \sigma
_3 , {\rd\over \rd x } \cdot \right]
\end{eqnarray}
whose spectrum is purely continuous.

On the other hand the expansions over the ``squared solutions''
can be understood as expansions over the eigenfunctions of
$\Lambda _+ $ and $\Lambda _- $. Note, that while $i d/dx $ does
not have discrete spectrum, the operators $\Lambda _\pm $ may have
discrete eigenvalues whenever $L $ has.  Finally, if we restrict
ourselves to the special class of reflectionless potentials for
which $\rho ^\pm(\lambda ) \equiv 0 $ for all $\lambda \in \bbbr
$, we obtain finite dimensional subspace of $\mathcal{ M} $. Then
the operators $\Lambda _\pm $ operate nontrivially only in this
subspace.

\section{Conclusions}
\label{sec:5}

We showed that the interpretation of the ISM as a generalized
Fourier transform holds true for the generalized Zakharov-Shabat
systems related to the symmetric spaces $SU(n+m)/S(U(n)\otimes
U(n)) $. Using the standard dressing method we outlined the
structure of the singularities of the fundamental analytic
solutions and constructed the soliton solutions with
multidimensional eigenspaces (the projectors $P_j(x) $ may have
rank $r_j>1 $).

The expansions over the `squared solutions' are natural tool to
derive the fundamental properties not only of the MNLS type
equations, but also of the NLEE with generic dispersion laws. Some
of these equations, besides the intriguing properties as dynamical
systems allowing for boomerons, trappons etc., may also have
interesting physical applications.

Another interesting area for further investigations is to study
and classify the reductions of these NLEE. For results along this
line for the MNLS equations see the reports \cite{varna04} and
\cite{manev04}; reductions of other types of NLEE have been
considered in \cite{G-cm,vgn,vgrn,LoMi1}.

One can also treat generalized Zakharov-Shabat systems related to
other symmetric spaces. This would require substantial changes in
the dressing factors. The expansions over the `squared solutions'
can be closely related to the graded Lie algebras, and to the
reduction group and provide an effective tool to derive and
analyze new soliton equations. For more details and further
reading see \cite{DrSok,G-cm}.

\section*{Acknowledgements.} This work has been supported by the
National Science Foundation of Bulgaria, contract No. F-1410.

\end{document}